\documentclass[
 reprint,
 amsmath,amssymb,
 aps,
 superscriptaddress,
 pre
]{revtex4-2}

\usepackage{xcolor}
\usepackage{graphicx}
\usepackage{dcolumn}
\usepackage{bm}
\usepackage{hyperref}
\hypersetup{colorlinks, linkcolor={blue},citecolor={blue}, urlcolor={blue}}

\begin{document}

\preprint{APS/123-QED}

\title{Implementing Non-Universal Features with a Random Matrix Theory Approach: \\Application to Space-to-Configuration Multiplexing}

\author{Philipp del Hougne}
\affiliation{Institut de Physique de Nice, CNRS UMR 7010, Universit\'{e} C\^ote d'Azur, 06108 Nice, France}

\author{Dmitry V. Savin}
\affiliation{Department of Mathematics, Brunel University London, Uxbridge, UB8 3PH, United Kingdom}%

\author{Olivier Legrand}
\affiliation{Institut de Physique de Nice, CNRS UMR 7010, Universit\'{e} C\^ote d'Azur, 06108 Nice, France}
\author{Ulrich Kuhl}
\affiliation{Institut de Physique de Nice, CNRS UMR 7010, Universit\'{e} C\^ote d'Azur, 06108 Nice, France}

\begin{abstract}

We consider the efficiency of multiplexing spatially encoded information across random configurations of a metasurface-programmable chaotic cavity in the microwave domain. The distribution of the effective rank of the channel matrix is studied to quantify the channel diversity and to assess a specific system’s performance. System-specific features such as unstirred field components give rise to nontrivial inter-channel correlations and need to be properly accounted for in modelling based on random matrix theory. To address this challenge, we propose a two-step hybrid approach. Based on an ensemble of experimentally measured scattering matrices for different random metasurface configurations, we first learn a system-specific pair of coupling matrix and unstirred contribution to the Hamiltonian, and then add an appropriately weighted stirred contribution. We verify that our method is capable of reproducing the experimentally found distribution of the effective rank with good accuracy. The approach can also be applied to other wave phenomena in complex media. 
\end{abstract}

\maketitle

The phenomenon of wave multiplexing underpins a wide range of applications in wave engineering~\cite{sebbah2001waves}. Sometimes, multiplexing devices are carefully engineered~\cite{lalau2013adjoint,molesky2018inverse,pande2020symphotic}; often, however, multiplexing is a result of the complete scrambling of wave fronts propagating through a complex medium (e.g., multiply scattering layer, chaotic cavity, multi-mode fiber). The most common example in the microwave domain is spatial multiplexing in multi-channel wireless communication due to an irregular propagation environment~\cite{telatar1999capacity,simon2001communication}. More recently, in order to circumvent the need for costly coherent receiver networks, the idea of multiplexing spatial information across different frequencies~\cite{hunt2013metamaterial,fromenteze2014waveform,fromenteze2015computational,fromenteze2016single} or configurations~\cite{sleasman2016microwave,asefi2017use,del2018dynamic,del2018precise,del2019optimized} of a complex medium gained traction in electromagnetic imaging and sensing.

Common to all multiplexing schemes is the description with the matrix formalism $Y=\mathcal{H}X$, linking the input vector $X$ to the output vector $Y$ via the channel matrix $\mathcal{H}$. The amount of correlations between different channels determines the quality of the multiplexing: the higher the correlations are, the more redundant information is acquired. Incidentally, this insight recently motivated efforts to tweak the disorder of complex media in order to reduce channel correlations via ``disorder-engineering'', both in space-to-space~\cite{del2019optimally} and space-to-configuration~\cite{del2019optimized} multiplexing. A convenient metric to quantify channel correlations is the effective rank, defined as $R_{\mathrm{eff}}(\mathcal{H}) = \mathrm{exp}\left( -\sum_{i=1}^n \tilde{\sigma}_i \mathrm{ln}(\tilde{\sigma}_i) \right)$, where $\tilde{\sigma}_i = \sigma_i / \sum_{i=1}^n \sigma_i$, $\sigma_i$ are the singular values of $\mathcal{H}$ and $n$ is the smaller one of the two dimensions of $\mathcal{H}$~\cite{roy2007effective}. Note that only perfectly orthogonal channels yield $R_{\mathrm{eff}}(\mathcal{H}) = n$.

The intuition used by the wave engineering community to interpret multiplexing phenomena and to conceive applications building upon them is largely based on the assumption that $\mathcal{H}$ resembles a random matrix with i.i.d.~Gaussian entries. Consider for concreteness space-to-configuration multiplexing with a reconfigurable chaotic cavity, as depicted in Fig.~1 and discussed in Refs.~\cite{sleasman2016microwave,asefi2017use,del2018dynamic,del2018precise,del2019optimized}. Ideally, such a system with considerable losses is indeed characterized by a perfectly stirred field distribution following the Rayleigh model~\cite{yeh2012first,kumar2013distribution}. 
In practice, however, a substantial unstirred field component persists, resulting in additional channel correlations and a lower-than-expected $R_{\mathrm{eff}}$~\cite{del2019optimized}. As a result, the wide-spread notion of degrees of freedom as corresponding to the dimensions of $\mathcal{H}$, rather than being related to its singular value spectrum~\cite{davy2012focusing,gradoni2019correlation,del2019optimized}, needs to be revisited~\cite{cozza2011skeptic,chen2013experimental}.

The unstirred field component also presents a significant challenge from a modelling point of view. Over the past decades, random matrix theory (RMT) has seen a large success in predicting universal features of a wide range of wave-chaotic systems~\cite{stockmann2000quantum,kuhl2005classical,fyodorov2005scattering,kuhl2013microwave}. In other words, non-universal features are usually removed before comparing experimental data to RMT predictions~\cite{engelbrecht1973hauser,hemmady2005universalPRL,hemmady2005universalPRE}. 
The crux of the space-to-configuration multiplexing system we consider lies, however, precisely in the non-universal features which cause additional correlations between channels. 
Our goal here is to model the statistical properties of our specific system's scattering matrix and to thereby reproduce the experimentally observed distribution of $R_{\mathrm{eff}}$.

System-specific features arise from reflections at the ports due to impedance mismatches~\cite{brouwer1995generalized,kuhl2005direct,fyodorov2004statistics}, from rays connecting two ports without ergodically exploring the cavity~\cite{baranger1996short,bulgakov2006statistical} and from ergodic rays that are not affected by the stirring mechanism. Refs.~\cite{hart2009effect,yeh2010universal,yeh2010experimental} report efforts to capture system-specific properties via semiclassical calculations of the average impedance matrix in terms of ray trajectories between ports. Such an analysis is not feasible for a geometrically complex three-dimensional cavity like the one we consider, see Fig.~1. 
An approach recently proposed in Refs.~\cite{savin2017fluctuations,savin2020statistics} addresses yet another system-specific mechanism: the presence of an established direct transmission mediated by a resonant mode coupled to the (isolated) complex environment. This approach is not applicable in our case, where scattering has a multimode nature with both modes and channels being coupled mainly to the same chaotic environment.

To overcome the above limitations, we propose a two-step approach to incorporate non-universal features in an RMT framework relying on an effective non-Hermitian Hamiltonian to describe the open wave system. First, we learn a system-specific coupling matrix. 
Second, we determine an appropriate weighting between the unstirred (deterministic) and stirred (chaotic) components of the Hamiltonian by employing a variant of parametric motion modelling~\cite{haake_chapter}. Rather than relying on knowing exact geometrical details of our system, we extract the scattering distributions of our system from experimental measurements of the scattering matrix for an ensemble of configurations. We find a good agreement between the distributions of $R_{\mathrm{eff}}$ observed in our experiments and RMT simulations. 

Our experiment, depicted in Fig.~1, consists of an irregularly shaped electrically large metallic cavity whose walls are partially covered with a reconfigurable reflect-array metasurface~\cite{del2019optimized}. At the working frequency of $5.6\ \mathrm{GHz}$, each of its 304 elements can be configured individually via the bias voltage of a PIN diode to mimic Dirichlet or Neumann boundary conditions~\cite{dupre2015wave}. By choosing different random configurations, the field inside the chaotic cavity can therefore be stirred in an all-electronic manner~\cite{sleasman2016microwave,del2018dynamic,del2018precise,del2019optimized,gros2019wave}. We modulate \textit{in situ} eight electromagnetic signals in amplitude and phase and inject them into the cavity via eight randomly located antennas. These eight pieces of information are multiplexed across $P$ configurations of the system by probing the field inside the cavity with a single antenna for $P$ random metasurface configurations. In total, $M=9$ antennas are thus connected to the cavity. For concreteness, we consider $P=8$ in this work. 
As illustrated in Fig.~1, each row of the resulting $8\times 8$ channel matrix $\mathcal{H}$ is part of a different system's $9 \times 9$ scattering matrix $S$. In our experiment, considering the ninth port as receiver, based on $500$ realizations we find the ensemble-averaged $\langle R_{\mathrm{eff}}\rangle = 4.11 \pm 0.20$ which is well below the value of $6.48 \pm 0.23$ expected for an $8 \times 8$ matrix whose random complex entries are i.i.d. Gaussian distributed~\cite{del2019optimized}.

\begin{figure} [t]
\centering
\includegraphics [width =  \columnwidth]{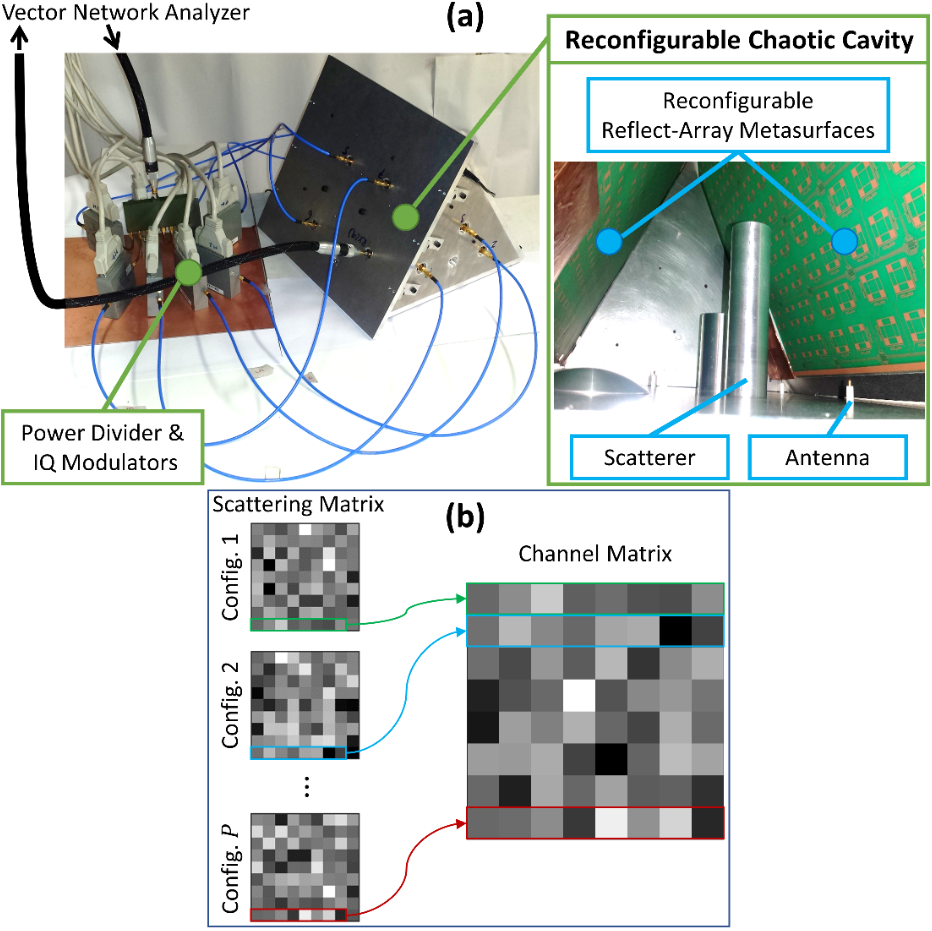}
\caption{(a) Experimental setup: Eight modulated signals are injected into a metasurface-reconfigurable chaotic cavity, the resulting field is probed with a ninth antenna. (b) The $P\times (M-1)$ channel matrix $\mathcal{H}$ is constructed for space-to-configuration multiplexing from entries of the $M \times M$ scattering matrix $S$ for $P$ different configurations of the system.}
\label{fig1}
\end{figure}

Following the RMT scattering approach~\cite{stockmann2000quantum,verbaarschot1985grassmann,mitchell2010random,dima_book_chapter}, we model our system as consisting of $N$ levels (modes) which are coupled to the environment via $M$ scattering channels. 
The coupling is described by an $N\times M$ real matrix $V$. The (energy-dependent) $M \times M$ scattering matrix $S^{RMT}(E)$ is represented in terms of the $N\times N$ Hamiltonian $H$ of the closed system and the coupling matrix $V$ as

\begin{equation}\label{Eq_S}
     S^{RMT}(E) = I_M - i V^T \frac{1}{ \left( E+i\frac{\Gamma_a}{2}\right)I_N - \left( H  -\frac{i}{2} V V^T \right) } V ,
\end{equation}

\noindent where $I_N$ denotes the $N \times N$ identity matrix and $\Gamma_a$ represents the dominant global absorption contribution to the resonance width. Conventionally, for an open chaotic system with time-reversal symmetry, $H$ is a real symmetric random matrix drawn from the Gaussian Orthogonal Ensemble (GOE) and the entries of $V$ are mutually independent zero-mean real Gaussian random variables. This ensures that $S^{RMT}$ is a symmetric matrix. We evaluate $S^{RMT}$ at $E=0$ and follow the normalization conventions used in Ref.~\cite{poli2009statistics}. 

To account for system-specific non-universal features, we alter two details in the above-described conventional RMT approach. First, we compose the Hamiltonian of an unstirred component $H^0$ and a stirred component $H^s$, weighted by a parameter $\lambda$:
\begin{equation}\label{Eq_H}
     H = \textrm{cos}(\lambda)H^0 + \textrm{sin}(\lambda) H^s.
\end{equation}
\noindent Both $H^0$ and $H^s$ are drawn from the GOE, but $H^0$ is kept fixed for different system realizations. The cosine and sine terms ensure that the Hamiltonian's eigenvalue probability density function (PDF) is not altered through our modification (see Sec.~11.10 in Ref.~\cite{haake_chapter}). Second, while $H^0$ is chosen randomly, we optimize $V$ and $\lambda$ such that the statistical properties of the resulting ensemble of $S^{RMT}$ match as closely as possible those of the experimental ensemble of $S^{EXP}$. 
We assume that the non-diagonal nature of the ensemble-averaged scattering matrix $\langle S \rangle$, i.e. the unstirred field components, are related to correlations between $V$ and $H^0$ in light of Eq.~(1); in principle, one could thus also keep $V$ fixed and optimize $H^0$~\footnote{Optimizing $H^0$ rather than $V$ has the inconvenience of resulting in $N^2+1$ as opposed to $NM+1$ parameters to be optimized.}. 

Our two-step optimization is based on the intuition that $V$ and $\lambda$ determine center and diameter, respectively, of the clouds of $S_{i,j}$ in the complex Argand diagram. We hypothesize that using $H=H^0$ or $H = \textrm{cos}(\lambda)H^0 + \textrm{sin}(\lambda) H^s$ in Eq.~(\ref{Eq_S}) will approximately yield the same $\langle S^{RMT} \rangle$ -- an intuitive assumption that we observe to be heuristically verified in Fig.~2(a). 
Therefore, using $H=H^0$ (and hence independent of $\lambda$), we can first optimize $V$ to ensure that $S^{RMT}_0 = \langle S^{EXP}\rangle$, where $S^{RMT}_0$ is the scattering matrix obtained via Eq.~(1) with $H=H^0$.
Once $V$ is fixed, we identify a suitable value of $\lambda$ to ensure that the standard deviations (SDs) of the $S_{i,j}$, a measure of the cloud sizes in the Argand diagram, match the experimentally observed ones.

\begin{figure} [t]
\centering
\includegraphics [width = \columnwidth]{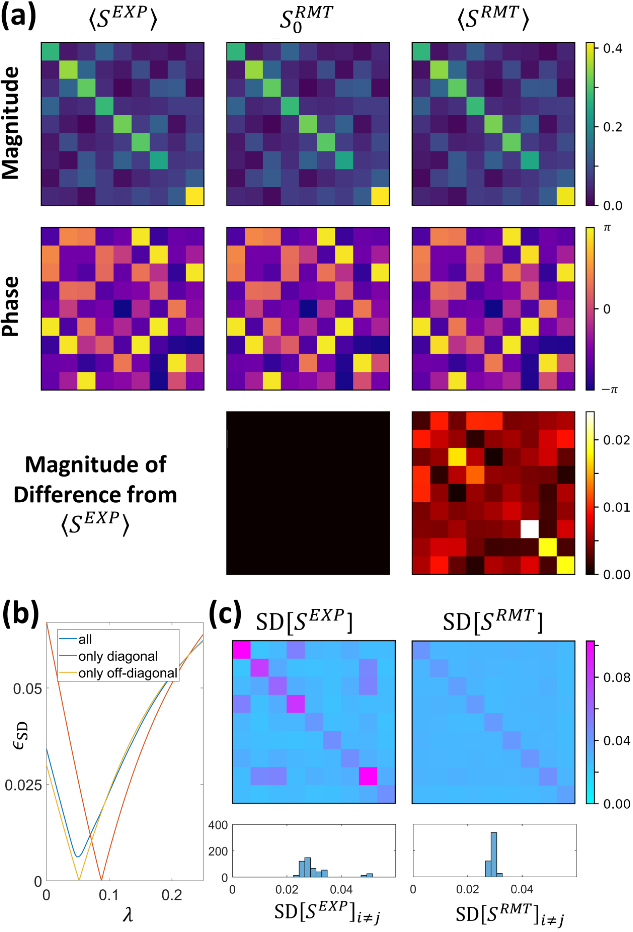}
\caption{(a) Comparison of $\langle S^{EXP} \rangle$, $S^{RMT}_0$ and $\langle S^{RMT} \rangle$ in terms of magnitude and phase. The magnitude of the difference of the latter two from $\langle S^{EXP} \rangle$ is also shown. (b) Dependence of $\epsilon_{SD}$ on $\lambda$ considering all entries, only diagonal entries or only off-diagonal entries of $S$. (c) SD of $S^{EXP}$ and $S^{RMT}$ and PDF of the SD of the off-diagonal entries of $S$.}
\label{fig2}
\end{figure}

We choose $N=250$ which is sufficiently large to ensure the local character of fluctuations but at the same time not too large in terms of the computational cost of optimizing $V$. 
Assuming that global absorption effects dominate, we extract $\Gamma_a = 7.4$ (in units of the mean level spacing) from the average decay rate of inverse Fourier transforms of experimentally measured transmission spectra. 
The optimization of $V$ is based on sequential least squares programming as originally introduced in Ref.~\cite{kraft1988software} and achieves equality of $S^{RMT}_0$ and $\langle S^{EXP}\rangle$ within large precision, as evidenced in Fig.~\ref{fig2}(a). The conventional RMT approach would assume that $\langle S \rangle$ is diagonal and $\mathrm{norm}(V_i)\approx \sqrt{2\kappa_i N/\pi}$, where $1-|\langle S_{ii} \rangle|^2 = 4\kappa_i / \left(1+\kappa_i\right)^2$. For our optimized coupling matrix, this is no longer the case in order to account for the above-mentioned correlations between $V$ and $H^0$. Moreover, we observe that for different optimization runs (for the same objective ensemble of $S^{EXP}$ but using a different $H^0$), completely different $\mathrm{norm}(V_i)$ are obtained. This is a typical observation in inverse problems, where usually a large number of different configurations (local optima) yields optimization outcomes of comparable quality.

Having optimized $V$, we now identify the optimal value of $\lambda$ based on the SDs of $S_{i,j}$ across $500$ realizations. For $SD[S^{EXP}]$, shown in Fig.~2(c), we notice that the average of the diagonal SDs (0.0673) is roughly double that of the off-diagonal SDs (0.0302). Moreover, the distribution of off-diagonal SDs shown in Fig.~\ref{fig2}(c) is narrowly peaked close to the average value but a weaker second peak at roughly double the value indicates that for a few entries of $S$ the SD is much stronger than on average. 
The fluctuations of the diagonal SDs for the experimental data are much larger than for the RMT data. 
To optimize $\lambda$, we search for the value that yields the lowest average error of the SDs of $S_{i,j}$, $\epsilon_{SD}$. As shown in Fig.~\ref{fig2}(b), $\lambda \approx 0.05$ is optimal if the error of the SD is averaged over all entries of $S$ or only over the off-diagonal entries of $S$. For the diagonal entries of $S$, however, $\lambda\approx0.09$ would be ideal. We use the former since our channel matrix is exclusively built from off-diagonal entries of $S$. 

Using the optimized $V$ and $\lambda$ in combination with an ensemble of 500 realizations of $H^s$ drawn from the GOE, we obtain an ensemble of 500 realizations of $S^{RMT}$ via Eq.~(1). As seen in Fig.~\ref{fig2}(a), its average value $\langle S^{RMT} \rangle$ is still extremely similar to $\langle S^{EXP} \rangle$, heuristically confirming our hypothesis that $\langle S^{RMT} \rangle$ is approximately independent of $\lambda$. The SDs obtained for the RMT ensemble, shown in Fig.~\ref{fig2}(c), nicely match the experimental ones for most off-diagonal entries. We notice that the RMT SDs are very uniform without any outliers, as evidenced by the narrow single-peaked PDF of the off-diagonal SDs. Consequently, our RMT model essentially appears to assume that all off-diagonal entries have the same SD of 0.03 and that all diagonal entries have a SD of 0.04. Different optimization runs starting with a different random $H^0$ yield very similar results.

Given our system-specific RMT model, we proceed with constructing the space-to-configuration multiplexing channel matrix $\mathcal{H}$ by picking eight random realizations out of the $500$ available ones. We repeat this $10^5$ times to compute the PDF of $R_{\mathrm{eff}}$. We perform this analysis for all nine possible choices of receiving antenna and for five optimization runs starting with a different random $H^0$. The results are summarized in Fig.~3.
To evidence how closely our proposed RMT model matches the experimental distribution in contrast to conventional RMT, we include the latter as benchmark in Fig.~3, too. Specifically, for each of the five runs with conventional RMT, we generate 500 scattering matrices by drawing 500 Hamiltonians from the GOE but keeping the random coupling vector $V$ fixed for all 500 realizations.
We also plot the PDF of $R_{\mathrm{eff}}$ obtained for an $8 \times 8$ matrix for which the real and imaginary components of the entries are simply drawn from a normal distribution. 

\begin{figure} [t]
\centering
\includegraphics [width = 7cm]{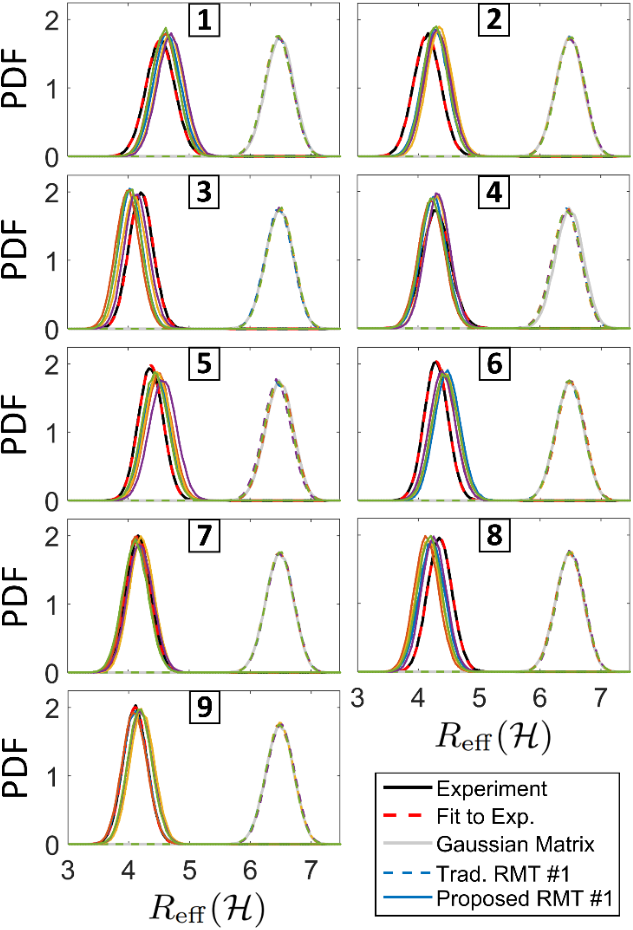}
\caption{For the nine different possible choices of receiving port (number indicated as inset), we compare the PDF of $R_{\mathrm{eff}}$ extracted from $S^{EXP}$ with that extract from $S^{RMT}$. For the latter, five different optimization runs (starting with a different random $H^0$) are shown in different colours (blue for run~\#1, etc.). A Gaussian fit to the experimental distribution is also shown. Moreover, we show the PDFs for five different ensembles with conventional RMT as well as for the case of $\mathcal{H}$ being a simple complex Gaussian matrix with i.i.d. entries.}
\label{fig3}
\end{figure}

It is evident in Fig.~3 that the PDFs of $R_{\mathrm{eff}}$ for conventional RMT and for such a Gaussian matrix are identical and independent of the choice of receiving port (i.e. of antenna coupling etc.). They clearly fail completely to predict the experimentally observed PDF, which motivates this paper. Moreover, it can be seen that the experimentally observed PDF is very well fitted with a Gaussian normal distribution. The PDFs resulting from five runs with our proposed RMT model differ slightly from run to run. Overall, the agreement of our proposed RMT model's PDF with the experimentally observed one is occasionally perfect (receiving port \#7) but always decent such that our proposed model serves its purpose.

\begin{figure} [t]
\centering
\includegraphics [width = \columnwidth]{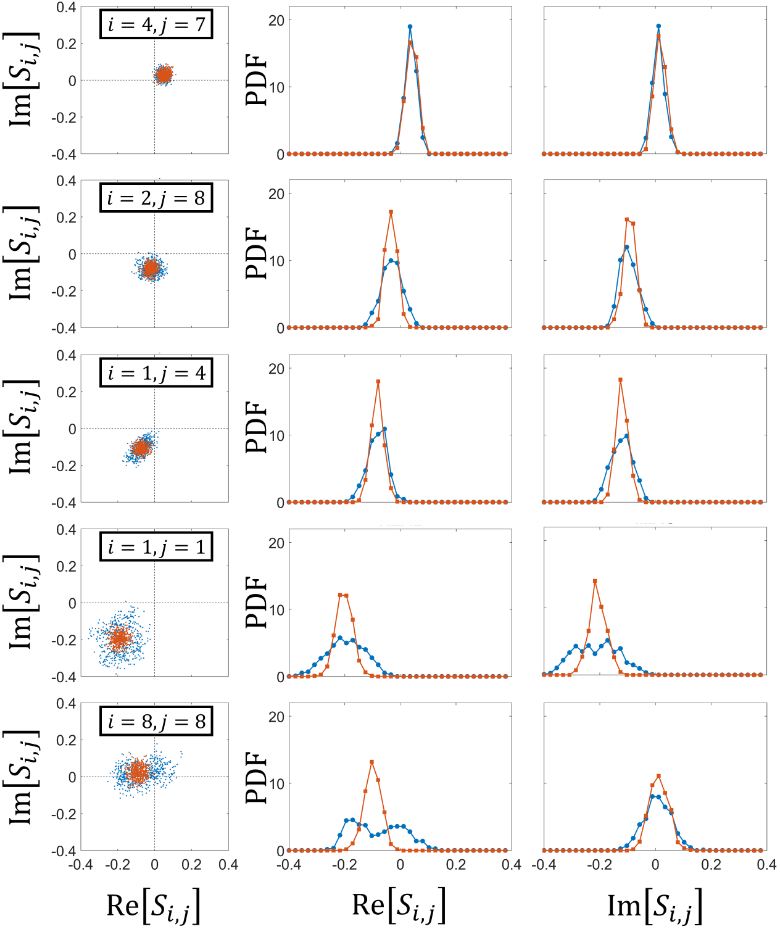}
\caption{Distribution of selected $S_{i,j}$ in the complex plane and corresponding PDFs of $\mathrm{Re}[S_{i,j}]$ and $\mathrm{Im}[S_{i,j}]$. Experiment and simulation are presented in blue and red, respectively.}
\label{fig4}
\end{figure}

Fig.~2(c) already hints at the reason why the agreement with the experiments is not perfect: the centered distributions of the $S_{i,j}$ are not identical for all $i,j$. To shed some light on this limitation of our model, we plot in Fig.~4 the distribution of selected $S_{i,j}$ in the complex Argand diagram as well as PDFs of real and imaginary component, using the data from the optimization run already considered in Fig.~\ref{fig2}. The first example, $S_{4,7}$, is representative of the behavior seen for most off-diagonal entries of $S$: the cloud in the complex plane is circular and the PDFs from our proposed RMT model match the experimentally obtained ones. One exception, illustrated by $S_{2,8}$, is due to clouds with a diameter that is significantly larger than the average. Another exception, namely $S_{1,4}$, is a case in which the cloud is not circular but elliptical. In both cases, our model fails to reproduce the PDFs correctly. For the off-diagonal entries of $S$, as already obvious in Fig.~2(c), our model's prediction is not very good. In general the diameter of the cloud is much larger in the experimental data, e.g. for $S_{1,1}$, and in some cases the cloud is not circular, as for $S_{8,8}$.

The $S_{i,j}$ clouds thus occasionally deviate from the expected behavior, either having a different radius or being deformed. Deformed (non-circular) clouds of $S_{i,j}$ were not seen in Ref.~\cite{gros2019wave}, possibly by chance because only a single $S_{i,j}$ was considered. Such exceptions imply that simply using centered quantities $S_{ij} - \langle S_{ij}\rangle$ does not always guarantee Rayleigh statistics in a tunable-metasurface-stirred chaotic cavity. 
Indeed, it was recently shown that in the presence of a deterministic scattering component the transmission amplitude and phase develop nontrivial statistical correlations even at strong absorption~\cite{savin2018envelope}.
Such correlations may also impact techniques using such centered quantities, for instance, for antenna characterization in reverberation chambers~\cite{holloway2012reverberation}.

Finally, we mention an alternative parametric model in which $H=H^0 + \lambda CC^\dagger$, where  $C$ is a $N \times N_{pix}$ matrix whose entries are zero-mean real Gaussian random variables and $N_{pix}$ is the (effective) number of metasurface pixels. 
This model could be motivated by a treatment of each metasurface element as an active scattering center or ``channel'' in one of two possible distinct reflecting states~\cite{conferenceJB}, which can be described by a low-rank perturbation of that type~\cite{kober2010microwave}. The weighted sum of a Gaussian Wigner and a Wishart matrix \cite{kumar2015random} brings about complications related to the distortion of the Hamiltonian's eigenvalue PDF. Our findings with this model, summarized in Ref.~\footnote{See the Supplemental Material for details on the alternative parametric model.}, show that overall it is not as compatible with our optimization protocol as is the model that we presented above. In particular, we saw that using $H=H^0$ or $H=H^0 + \lambda CC^\dagger$ does not yield the same $\langle S \rangle$ (within reasonable precision), a crucial assumption for the applicability of our two-step optimization procedure.

To summarize, we have introduced a modified RMT framework to capture system-specific features based on the measurements of an ensemble of realizations of the experimental scattering matrix. Our approach, first, optimizes the coupling matrix to ensure the correct $\langle S \rangle$ is obtained and, second, adds an appropriately weighted stirred contribution to the Hamiltonian to adjust the fluctuations of the scattering parameters. We found a good agreement with the experimentally obtained distribution of $R_{\mathrm{eff}}$ that characterizes our space-to-configuration multiplexing system. 
Our modified RMT scheme can also be applied to other scenarios where an ensemble of realizations of a chaotic system contains non-universal features, for instance, in the context of antenna characterization in stirred reverberation chambers~\cite{holloway2012reverberation,davy2016green}.

\smallskip

P.d.H., O.L. and U.K. are supported by the French “Agence Nationale de la Recherche” under reference ANR-17-ASTR-0017. The metasurface prototypes were purchased from Greenerwave. The authors acknowledge fruitful discussions with Fabrice Mortessagne.

\providecommand{\noopsort}[1]{}\providecommand{\singleletter}[1]{#1}%


\begin{thebibliography}{56}%
\makeatletter
\providecommand \@ifxundefined [1]{%
 \@ifx{#1\undefined}
}%
\providecommand \@ifnum [1]{%
 \ifnum #1\expandafter \@firstoftwo
 \else \expandafter \@secondoftwo
 \fi
}%
\providecommand \@ifx [1]{%
 \ifx #1\expandafter \@firstoftwo
 \else \expandafter \@secondoftwo
 \fi
}%
\providecommand \natexlab [1]{#1}%
\providecommand \enquote  [1]{``#1''}%
\providecommand \bibnamefont  [1]{#1}%
\providecommand \bibfnamefont [1]{#1}%
\providecommand \citenamefont [1]{#1}%
\providecommand \href@noop [0]{\@secondoftwo}%
\providecommand \href [0]{\begingroup \@sanitize@url \@href}%
\providecommand \@href[1]{\@@startlink{#1}\@@href}%
\providecommand \@@href[1]{\endgroup#1\@@endlink}%
\providecommand \@sanitize@url [0]{\catcode `\\12\catcode `\$12\catcode
  `\&12\catcode `\#12\catcode `\^12\catcode `\_12\catcode `\%12\relax}%
\providecommand \@@startlink[1]{}%
\providecommand \@@endlink[0]{}%
\providecommand \url  [0]{\begingroup\@sanitize@url \@url }%
\providecommand \@url [1]{\endgroup\@href {#1}{\urlprefix }}%
\providecommand \urlprefix  [0]{URL }%
\providecommand \Eprint [0]{\href }%
\providecommand \doibase [0]{https://doi.org/}%
\providecommand \selectlanguage [0]{\@gobble}%
\providecommand \bibinfo  [0]{\@secondoftwo}%
\providecommand \bibfield  [0]{\@secondoftwo}%
\providecommand \translation [1]{[#1]}%
\providecommand \BibitemOpen [0]{}%
\providecommand \bibitemStop [0]{}%
\providecommand \bibitemNoStop [0]{.\EOS\space}%
\providecommand \EOS [0]{\spacefactor3000\relax}%
\providecommand \BibitemShut  [1]{\csname bibitem#1\endcsname}%
\let\auto@bib@innerbib\@empty
%</preamble>
\bibitem [{\citenamefont {Sebbah}(2001)}]{sebbah2001waves}%
  \BibitemOpen
  \bibfield  {author} {\bibinfo {author} {\bibfnamefont {P.}~\bibnamefont
  {Sebbah}},\ }\href@noop {} {\emph {\bibinfo {title} {Waves and Imaging
  through Complex Media}}}\ (\bibinfo  {publisher} {Springer Science \&
  Business Media},\ \bibinfo {year} {2001})\BibitemShut {NoStop}%
\bibitem [{\citenamefont {Lalau-Keraly}\ \emph {et~al.}(2013)\citenamefont
  {Lalau-Keraly}, \citenamefont {Bhargava}, \citenamefont {Miller},\ and\
  \citenamefont {Yablonovitch}}]{lalau2013adjoint}%
  \BibitemOpen
  \bibfield  {author} {\bibinfo {author} {\bibfnamefont {C.~M.}\ \bibnamefont
  {Lalau-Keraly}}, \bibinfo {author} {\bibfnamefont {S.}~\bibnamefont
  {Bhargava}}, \bibinfo {author} {\bibfnamefont {O.~D.}\ \bibnamefont
  {Miller}},\ and\ \bibinfo {author} {\bibfnamefont {E.}~\bibnamefont
  {Yablonovitch}},\ }\bibfield  {title} {\bibinfo {title} {Adjoint shape
  optimization applied to electromagnetic design},\ }\href
  {https://doi.org/10.1364/OE.21.021693} {\bibfield  {journal} {\bibinfo
  {journal} {Opt. Express}\ }\textbf {\bibinfo {volume} {21}},\ \bibinfo
  {pages} {21693} (\bibinfo {year} {2013})}\BibitemShut {NoStop}%
\bibitem [{\citenamefont {Molesky}\ \emph {et~al.}(2018)\citenamefont
  {Molesky}, \citenamefont {Lin}, \citenamefont {Piggott}, \citenamefont {Jin},
  \citenamefont {Vuckovi{\'c}},\ and\ \citenamefont
  {Rodriguez}}]{molesky2018inverse}%
  \BibitemOpen
  \bibfield  {author} {\bibinfo {author} {\bibfnamefont {S.}~\bibnamefont
  {Molesky}}, \bibinfo {author} {\bibfnamefont {Z.}~\bibnamefont {Lin}},
  \bibinfo {author} {\bibfnamefont {A.~Y.}\ \bibnamefont {Piggott}}, \bibinfo
  {author} {\bibfnamefont {W.}~\bibnamefont {Jin}}, \bibinfo {author}
  {\bibfnamefont {J.}~\bibnamefont {Vuckovi{\'c}}},\ and\ \bibinfo {author}
  {\bibfnamefont {A.~W.}\ \bibnamefont {Rodriguez}},\ }\bibfield  {title}
  {\bibinfo {title} {Inverse design in nanophotonics},\ }\href
  {https://doi.org/10.1038/s41566-018-0246-9} {\bibfield  {journal} {\bibinfo
  {journal} {Nat. Photonics}\ }\textbf {\bibinfo {volume} {12}},\ \bibinfo
  {pages} {659} (\bibinfo {year} {2018})}\BibitemShut {NoStop}%
\bibitem [{\citenamefont {Pande}\ \emph {et~al.}(2020)\citenamefont {Pande},
  \citenamefont {Gollub}, \citenamefont {Zecca}, \citenamefont {Marks},\ and\
  \citenamefont {Smith}}]{pande2020symphotic}%
  \BibitemOpen
  \bibfield  {author} {\bibinfo {author} {\bibfnamefont {D.}~\bibnamefont
  {Pande}}, \bibinfo {author} {\bibfnamefont {J.}~\bibnamefont {Gollub}},
  \bibinfo {author} {\bibfnamefont {R.}~\bibnamefont {Zecca}}, \bibinfo
  {author} {\bibfnamefont {D.~L.}\ \bibnamefont {Marks}},\ and\ \bibinfo
  {author} {\bibfnamefont {D.~R.}\ \bibnamefont {Smith}},\ }\bibfield  {title}
  {\bibinfo {title} {Symphotic multiplexing medium at microwave frequencies},\
  }\href {https://doi.org/10.1103/PhysRevApplied.13.024033} {\bibfield
  {journal} {\bibinfo  {journal} {Phys. Rev. Applied}\ }\textbf {\bibinfo
  {volume} {13}},\ \bibinfo {pages} {024033} (\bibinfo {year}
  {2020})}\BibitemShut {NoStop}%
\bibitem [{\citenamefont {Telatar}(1999)}]{telatar1999capacity}%
  \BibitemOpen
  \bibfield  {author} {\bibinfo {author} {\bibfnamefont {E.}~\bibnamefont
  {Telatar}},\ }\bibfield  {title} {\bibinfo {title} {{Capacity of
  Multiâ€antenna Gaussian Channels}},\ }\href
  {https://doi.org/10.1002/ett.4460100604} {\bibfield  {journal} {\bibinfo
  {journal} {Eur. Trans. Telecomm.}\ }\textbf {\bibinfo {volume} {10}},\
  \bibinfo {pages} {585} (\bibinfo {year} {1999})}\BibitemShut {NoStop}%
\bibitem [{\citenamefont {Simon}\ \emph {et~al.}(2001)\citenamefont {Simon},
  \citenamefont {Moustakas}, \citenamefont {Stoytchev},\ and\ \citenamefont
  {Safar}}]{simon2001communication}%
  \BibitemOpen
  \bibfield  {author} {\bibinfo {author} {\bibfnamefont {S.~H.}\ \bibnamefont
  {Simon}}, \bibinfo {author} {\bibfnamefont {A.~L.}\ \bibnamefont
  {Moustakas}}, \bibinfo {author} {\bibfnamefont {M.}~\bibnamefont
  {Stoytchev}},\ and\ \bibinfo {author} {\bibfnamefont {H.}~\bibnamefont
  {Safar}},\ }\bibfield  {title} {\bibinfo {title} {Communication in a
  disordered world},\ }\href {https://doi.org/10.1063/1.1420510} {\bibfield
  {journal} {\bibinfo  {journal} {Phys. Today}\ }\textbf {\bibinfo {volume}
  {54}} (\bibinfo {year} {2001})}\BibitemShut {NoStop}%
\bibitem [{\citenamefont {Hunt}\ \emph {et~al.}(2013)\citenamefont {Hunt},
  \citenamefont {Driscoll}, \citenamefont {Mrozack}, \citenamefont {Lipworth},
  \citenamefont {Reynolds}, \citenamefont {Brady},\ and\ \citenamefont
  {Smith}}]{hunt2013metamaterial}%
  \BibitemOpen
  \bibfield  {author} {\bibinfo {author} {\bibfnamefont {J.}~\bibnamefont
  {Hunt}}, \bibinfo {author} {\bibfnamefont {T.}~\bibnamefont {Driscoll}},
  \bibinfo {author} {\bibfnamefont {A.}~\bibnamefont {Mrozack}}, \bibinfo
  {author} {\bibfnamefont {G.}~\bibnamefont {Lipworth}}, \bibinfo {author}
  {\bibfnamefont {M.}~\bibnamefont {Reynolds}}, \bibinfo {author}
  {\bibfnamefont {D.}~\bibnamefont {Brady}},\ and\ \bibinfo {author}
  {\bibfnamefont {D.~R.}\ \bibnamefont {Smith}},\ }\bibfield  {title} {\bibinfo
  {title} {{Metamaterial Apertures for Computational Imaging}},\ }\href
  {https://doi.org/10.1126/science.1230054} {\bibfield  {journal} {\bibinfo
  {journal} {Science}\ }\textbf {\bibinfo {volume} {339}},\ \bibinfo {pages}
  {310} (\bibinfo {year} {2013})}\BibitemShut {NoStop}%
\bibitem [{\citenamefont {Fromenteze}\ \emph {et~al.}(2014)\citenamefont
  {Fromenteze}, \citenamefont {Decroze},\ and\ \citenamefont
  {Carsenat}}]{fromenteze2014waveform}%
  \BibitemOpen
  \bibfield  {author} {\bibinfo {author} {\bibfnamefont {T.}~\bibnamefont
  {Fromenteze}}, \bibinfo {author} {\bibfnamefont {C.}~\bibnamefont
  {Decroze}},\ and\ \bibinfo {author} {\bibfnamefont {D.}~\bibnamefont
  {Carsenat}},\ }\bibfield  {title} {\bibinfo {title} {{Waveform Coding for
  Passive Multiplexing: Application to Microwave Imaging}},\ }\href
  {https://doi.org/10.1109/TAP.2014.2382647} {\bibfield  {journal} {\bibinfo
  {journal} {IEEE Trans. Antennas Propag.}\ }\textbf {\bibinfo {volume} {63}},\
  \bibinfo {pages} {593} (\bibinfo {year} {2014})}\BibitemShut {NoStop}%
\bibitem [{\citenamefont {Fromenteze}\ \emph {et~al.}(2015)\citenamefont
  {Fromenteze}, \citenamefont {Yurduseven}, \citenamefont {Imani},
  \citenamefont {Gollub}, \citenamefont {Decroze}, \citenamefont {Carsenat},\
  and\ \citenamefont {Smith}}]{fromenteze2015computational}%
  \BibitemOpen
  \bibfield  {author} {\bibinfo {author} {\bibfnamefont {T.}~\bibnamefont
  {Fromenteze}}, \bibinfo {author} {\bibfnamefont {O.}~\bibnamefont
  {Yurduseven}}, \bibinfo {author} {\bibfnamefont {M.~F.}\ \bibnamefont
  {Imani}}, \bibinfo {author} {\bibfnamefont {J.}~\bibnamefont {Gollub}},
  \bibinfo {author} {\bibfnamefont {C.}~\bibnamefont {Decroze}}, \bibinfo
  {author} {\bibfnamefont {D.}~\bibnamefont {Carsenat}},\ and\ \bibinfo
  {author} {\bibfnamefont {D.~R.}\ \bibnamefont {Smith}},\ }\bibfield  {title}
  {\bibinfo {title} {Computational imaging using a mode-mixing cavity at
  microwave frequencies},\ }\href {https://doi.org/10.1063/1.4921081}
  {\bibfield  {journal} {\bibinfo  {journal} {Appl. Phys. Lett.}\ }\textbf
  {\bibinfo {volume} {106}},\ \bibinfo {pages} {194104} (\bibinfo {year}
  {2015})}\BibitemShut {NoStop}%
\bibitem [{\citenamefont {Fromenteze}\ \emph {et~al.}(2016)\citenamefont
  {Fromenteze}, \citenamefont {Kpr{\'e}}, \citenamefont {Carsenat},
  \citenamefont {Decroze},\ and\ \citenamefont
  {Sakamoto}}]{fromenteze2016single}%
  \BibitemOpen
  \bibfield  {author} {\bibinfo {author} {\bibfnamefont {T.}~\bibnamefont
  {Fromenteze}}, \bibinfo {author} {\bibfnamefont {E.~L.}\ \bibnamefont
  {Kpr{\'e}}}, \bibinfo {author} {\bibfnamefont {D.}~\bibnamefont {Carsenat}},
  \bibinfo {author} {\bibfnamefont {C.}~\bibnamefont {Decroze}},\ and\ \bibinfo
  {author} {\bibfnamefont {T.}~\bibnamefont {Sakamoto}},\ }\bibfield  {title}
  {\bibinfo {title} {{Single-Shot Compressive Multiple-Inputs Multiple-Outputs
  Radar Imaging Using a Two-Port Passive Device}},\ }\href
  {https://doi.org/10.1109/ACCESS.2016.2543525} {\bibfield  {journal} {\bibinfo
   {journal} {IEEE Access}\ }\textbf {\bibinfo {volume} {4}},\ \bibinfo {pages}
  {1050} (\bibinfo {year} {2016})}\BibitemShut {NoStop}%
\bibitem [{\citenamefont {Sleasman}\ \emph {et~al.}(2016)\citenamefont
  {Sleasman}, \citenamefont {Imani}, \citenamefont {Gollub},\ and\
  \citenamefont {Smith}}]{sleasman2016microwave}%
  \BibitemOpen
  \bibfield  {author} {\bibinfo {author} {\bibfnamefont {T.}~\bibnamefont
  {Sleasman}}, \bibinfo {author} {\bibfnamefont {M.~F.}\ \bibnamefont {Imani}},
  \bibinfo {author} {\bibfnamefont {J.~N.}\ \bibnamefont {Gollub}},\ and\
  \bibinfo {author} {\bibfnamefont {D.~R.}\ \bibnamefont {Smith}},\ }\bibfield
  {title} {\bibinfo {title} {{Microwave Imaging Using a Disordered Cavity with
  a Dynamically Tunable Impedance Surface}},\ }\href
  {https://doi.org/10.1103/PhysRevApplied.6.054019} {\bibfield  {journal}
  {\bibinfo  {journal} {Phys. Rev. Applied}\ }\textbf {\bibinfo {volume} {6}},\
  \bibinfo {pages} {054019} (\bibinfo {year} {2016})}\BibitemShut {NoStop}%
\bibitem [{\citenamefont {Asefi}\ and\ \citenamefont
  {LoVetri}(2017)}]{asefi2017use}%
  \BibitemOpen
  \bibfield  {author} {\bibinfo {author} {\bibfnamefont {M.}~\bibnamefont
  {Asefi}}\ and\ \bibinfo {author} {\bibfnamefont {J.}~\bibnamefont
  {LoVetri}},\ }\bibfield  {title} {\bibinfo {title} {{Use of Field-Perturbing
  Elements to Increase Nonredundant Data for Microwave Imaging Systems}},\
  }\href {https://doi.org/10.1109/TMTT.2017.2681657} {\bibfield  {journal}
  {\bibinfo  {journal} {IEEE Trans. Microw. Theory Tech.}\ }\textbf {\bibinfo
  {volume} {65}},\ \bibinfo {pages} {3172} (\bibinfo {year}
  {2017})}\BibitemShut {NoStop}%
\bibitem [{\citenamefont {del Hougne}\ \emph
  {et~al.}(2018{\natexlab{a}})\citenamefont {del Hougne}, \citenamefont
  {Imani}, \citenamefont {Sleasman}, \citenamefont {Gollub}, \citenamefont
  {Fink}, \citenamefont {Lerosey},\ and\ \citenamefont
  {Smith}}]{del2018dynamic}%
  \BibitemOpen
  \bibfield  {author} {\bibinfo {author} {\bibfnamefont {P.}~\bibnamefont {del
  Hougne}}, \bibinfo {author} {\bibfnamefont {M.~F.}\ \bibnamefont {Imani}},
  \bibinfo {author} {\bibfnamefont {T.}~\bibnamefont {Sleasman}}, \bibinfo
  {author} {\bibfnamefont {J.~N.}\ \bibnamefont {Gollub}}, \bibinfo {author}
  {\bibfnamefont {M.}~\bibnamefont {Fink}}, \bibinfo {author} {\bibfnamefont
  {G.}~\bibnamefont {Lerosey}},\ and\ \bibinfo {author} {\bibfnamefont {D.~R.}\
  \bibnamefont {Smith}},\ }\bibfield  {title} {\bibinfo {title} {Dynamic
  metasurface aperture as smart around-the-corner motion detector},\ }\href
  {https://doi.org/10.1038/s41598-018-24681-9} {\bibfield  {journal} {\bibinfo
  {journal} {Sci. Rep.}\ }\textbf {\bibinfo {volume} {8}},\ \bibinfo {pages}
  {1} (\bibinfo {year} {2018}{\natexlab{a}})}\BibitemShut {NoStop}%
\bibitem [{\citenamefont {del Hougne}\ \emph
  {et~al.}(2018{\natexlab{b}})\citenamefont {del Hougne}, \citenamefont
  {Imani}, \citenamefont {Fink}, \citenamefont {Smith},\ and\ \citenamefont
  {Lerosey}}]{del2018precise}%
  \BibitemOpen
  \bibfield  {author} {\bibinfo {author} {\bibfnamefont {P.}~\bibnamefont {del
  Hougne}}, \bibinfo {author} {\bibfnamefont {M.~F.}\ \bibnamefont {Imani}},
  \bibinfo {author} {\bibfnamefont {M.}~\bibnamefont {Fink}}, \bibinfo {author}
  {\bibfnamefont {D.~R.}\ \bibnamefont {Smith}},\ and\ \bibinfo {author}
  {\bibfnamefont {G.}~\bibnamefont {Lerosey}},\ }\bibfield  {title} {\bibinfo
  {title} {{Precise Localization of Multiple Noncooperative Objects in a
  Disordered Cavity by Wave Front Shaping}},\ }\href
  {https://doi.org/10.1103/PhysRevLett.121.063901} {\bibfield  {journal}
  {\bibinfo  {journal} {Phys. Rev. Lett.}\ }\textbf {\bibinfo {volume} {121}},\
  \bibinfo {pages} {063901} (\bibinfo {year} {2018}{\natexlab{b}})}\BibitemShut
  {NoStop}%
\bibitem [{\citenamefont {del Hougne}\ \emph {et~al.}(2020)\citenamefont {del
  Hougne}, \citenamefont {Davy},\ and\ \citenamefont
  {Kuhl}}]{del2019optimized}%
  \BibitemOpen
  \bibfield  {author} {\bibinfo {author} {\bibfnamefont {P.}~\bibnamefont {del
  Hougne}}, \bibinfo {author} {\bibfnamefont {M.}~\bibnamefont {Davy}},\ and\
  \bibinfo {author} {\bibfnamefont {U.}~\bibnamefont {Kuhl}},\ }\bibfield
  {title} {\bibinfo {title} {{Optimal Multiplexing of Spatially Encoded
  Information across Custom-Tailored Configurations of a Metasurface-Tunable
  Chaotic Cavity}},\ }\href {https://doi.org/10.1103/PhysRevApplied.13.041004}
  {\bibfield  {journal} {\bibinfo  {journal} {Phys. Rev. Applied}\ }\textbf
  {\bibinfo {volume} {13}},\ \bibinfo {pages} {041004} (\bibinfo {year}
  {2020})}\BibitemShut {NoStop}%
\bibitem [{\citenamefont {del Hougne}\ \emph {et~al.}(2019)\citenamefont {del
  Hougne}, \citenamefont {Fink},\ and\ \citenamefont
  {Lerosey}}]{del2019optimally}%
  \BibitemOpen
  \bibfield  {author} {\bibinfo {author} {\bibfnamefont {P.}~\bibnamefont {del
  Hougne}}, \bibinfo {author} {\bibfnamefont {M.}~\bibnamefont {Fink}},\ and\
  \bibinfo {author} {\bibfnamefont {G.}~\bibnamefont {Lerosey}},\ }\bibfield
  {title} {\bibinfo {title} {Optimally diverse communication channels in
  disordered environments with tuned randomness},\ }\href
  {https://doi.org/10.1038/s41928-018-0190-1} {\bibfield  {journal} {\bibinfo
  {journal} {Nat. Electron.}\ }\textbf {\bibinfo {volume} {2}},\ \bibinfo
  {pages} {36} (\bibinfo {year} {2019})}\BibitemShut {NoStop}%
\bibitem [{\citenamefont {Roy}\ and\ \citenamefont
  {Vetterli}(2007)}]{roy2007effective}%
  \BibitemOpen
  \bibfield  {author} {\bibinfo {author} {\bibfnamefont {O.}~\bibnamefont
  {Roy}}\ and\ \bibinfo {author} {\bibfnamefont {M.}~\bibnamefont {Vetterli}},\
  }\bibfield  {title} {\bibinfo {title} {{The effective rank: A measure of
  effective dimensionality}},\ }in\ \href
  {https://ieeexplore.ieee.org/abstract/document/7098875} {\emph {\bibinfo
  {booktitle} {15th European Signal Processing Conference}}}\ (\bibinfo
  {organization} {IEEE},\ \bibinfo {year} {2007})\ pp.\ \bibinfo {pages}
  {606--610}\BibitemShut {NoStop}%
\bibitem [{\citenamefont {Yeh}\ \emph {et~al.}(2012)\citenamefont {Yeh},
  \citenamefont {Antonsen}, \citenamefont {Ott},\ and\ \citenamefont
  {Anlage}}]{yeh2012first}%
  \BibitemOpen
  \bibfield  {author} {\bibinfo {author} {\bibfnamefont {J.-H.}\ \bibnamefont
  {Yeh}}, \bibinfo {author} {\bibfnamefont {T.~M.}\ \bibnamefont {Antonsen}},
  \bibinfo {author} {\bibfnamefont {E.}~\bibnamefont {Ott}},\ and\ \bibinfo
  {author} {\bibfnamefont {S.~M.}\ \bibnamefont {Anlage}},\ }\bibfield  {title}
  {\bibinfo {title} {First-principles model of time-dependent variations in
  transmission through a fluctuating scattering environment},\ }\href
  {https://doi.org/10.1103/PhysRevE.85.015202} {\bibfield  {journal} {\bibinfo
  {journal} {Phys. Rev. E}\ }\textbf {\bibinfo {volume} {85}},\ \bibinfo
  {pages} {015202} (\bibinfo {year} {2012})}\BibitemShut {NoStop}%
\bibitem [{\citenamefont {Kumar}\ \emph {et~al.}(2013)\citenamefont {Kumar},
  \citenamefont {Nock}, \citenamefont {Sommers}, \citenamefont {Guhr},
  \citenamefont {Dietz}, \citenamefont {Miski-Oglu}, \citenamefont {Richter},\
  and\ \citenamefont {Sch{\"a}fer}}]{kumar2013distribution}%
  \BibitemOpen
  \bibfield  {author} {\bibinfo {author} {\bibfnamefont {S.}~\bibnamefont
  {Kumar}}, \bibinfo {author} {\bibfnamefont {A.}~\bibnamefont {Nock}},
  \bibinfo {author} {\bibfnamefont {H.-J.}\ \bibnamefont {Sommers}}, \bibinfo
  {author} {\bibfnamefont {T.}~\bibnamefont {Guhr}}, \bibinfo {author}
  {\bibfnamefont {B.}~\bibnamefont {Dietz}}, \bibinfo {author} {\bibfnamefont
  {M.}~\bibnamefont {Miski-Oglu}}, \bibinfo {author} {\bibfnamefont
  {A.}~\bibnamefont {Richter}},\ and\ \bibinfo {author} {\bibfnamefont
  {F.}~\bibnamefont {Sch{\"a}fer}},\ }\bibfield  {title} {\bibinfo {title}
  {{Distribution of Scattering Matrix Elements in Quantum Chaotic
  Scattering}},\ }\href {https://doi.org/10.1103/PhysRevLett.111.030403}
  {\bibfield  {journal} {\bibinfo  {journal} {Phys. Rev. Lett.}\ }\textbf
  {\bibinfo {volume} {111}},\ \bibinfo {pages} {030403} (\bibinfo {year}
  {2013})}\BibitemShut {NoStop}%
\bibitem [{\citenamefont {Davy}\ \emph {et~al.}(2012)\citenamefont {Davy},
  \citenamefont {Shi},\ and\ \citenamefont {Genack}}]{davy2012focusing}%
  \BibitemOpen
  \bibfield  {author} {\bibinfo {author} {\bibfnamefont {M.}~\bibnamefont
  {Davy}}, \bibinfo {author} {\bibfnamefont {Z.}~\bibnamefont {Shi}},\ and\
  \bibinfo {author} {\bibfnamefont {A.~Z.}\ \bibnamefont {Genack}},\ }\bibfield
   {title} {\bibinfo {title} {Focusing through random media: Eigenchannel
  participation number and intensity correlation},\ }\href
  {https://doi.org/10.1103/PhysRevB.85.035105} {\bibfield  {journal} {\bibinfo
  {journal} {Phys. Rev. B}\ }\textbf {\bibinfo {volume} {85}},\ \bibinfo
  {pages} {035105} (\bibinfo {year} {2012})}\BibitemShut {NoStop}%
\bibitem [{\citenamefont {Gradoni}\ \emph {et~al.}(2019)\citenamefont
  {Gradoni}, \citenamefont {Moglie},\ and\ \citenamefont
  {Primiani}}]{gradoni2019correlation}%
  \BibitemOpen
  \bibfield  {author} {\bibinfo {author} {\bibfnamefont {G.}~\bibnamefont
  {Gradoni}}, \bibinfo {author} {\bibfnamefont {F.}~\bibnamefont {Moglie}},\
  and\ \bibinfo {author} {\bibfnamefont {V.~M.}\ \bibnamefont {Primiani}},\
  }\bibfield  {title} {\bibinfo {title} {Correlation matrix methods to assess
  the stirring performance of electromagnetic reverberation chambers},\ }\href
  {https://doi.org/10.1016/j.wavemoti.2018.09.008} {\bibfield  {journal}
  {\bibinfo  {journal} {Wave Motion}\ }\textbf {\bibinfo {volume} {87}},\
  \bibinfo {pages} {213} (\bibinfo {year} {2019})}\BibitemShut {NoStop}%
\bibitem [{\citenamefont {Cozza}(2011)}]{cozza2011skeptic}%
  \BibitemOpen
  \bibfield  {author} {\bibinfo {author} {\bibfnamefont {A.}~\bibnamefont
  {Cozza}},\ }\bibfield  {title} {\bibinfo {title} {A skeptic's view of
  unstirred components},\ }in\ \href
  {https://ieeexplore.ieee.org/abstract/document/6078535} {\emph {\bibinfo
  {booktitle} {10th International Symposium on Electromagnetic
  Compatibility}}}\ (\bibinfo {organization} {IEEE},\ \bibinfo {year} {2011})\
  pp.\ \bibinfo {pages} {174--179}\BibitemShut {NoStop}%
\bibitem [{\citenamefont {Chen}(2013)}]{chen2013experimental}%
  \BibitemOpen
  \bibfield  {author} {\bibinfo {author} {\bibfnamefont {X.}~\bibnamefont
  {Chen}},\ }\bibfield  {title} {\bibinfo {title} {Experimental investigation
  of the number of independent samples and the measurement uncertainty in a
  reverberation chamber},\ }\href {https://doi.org/10.1109/TEMC.2013.2242473}
  {\bibfield  {journal} {\bibinfo  {journal} {IEEE Trans. Electromagn.
  Compat.}\ }\textbf {\bibinfo {volume} {55}},\ \bibinfo {pages} {816}
  (\bibinfo {year} {2013})}\BibitemShut {NoStop}%
\bibitem [{\citenamefont {St{\"o}ckmann}(2000)}]{stockmann2000quantum}%
  \BibitemOpen
  \bibfield  {author} {\bibinfo {author} {\bibfnamefont {H.-J.}\ \bibnamefont
  {St{\"o}ckmann}},\ }\href@noop {} {\bibinfo {title} {{Quantum Chaos: An
  Introduction}}} (\bibinfo {year} {2000})\BibitemShut {NoStop}%
\bibitem [{\citenamefont {Kuhl}\ \emph
  {et~al.}(2005{\natexlab{a}})\citenamefont {Kuhl}, \citenamefont
  {St{\"o}ckmann},\ and\ \citenamefont {Weaver}}]{kuhl2005classical}%
  \BibitemOpen
  \bibfield  {author} {\bibinfo {author} {\bibfnamefont {U.}~\bibnamefont
  {Kuhl}}, \bibinfo {author} {\bibfnamefont {H.}~\bibnamefont
  {St{\"o}ckmann}},\ and\ \bibinfo {author} {\bibfnamefont {R.}~\bibnamefont
  {Weaver}},\ }\bibfield  {title} {\bibinfo {title} {Classical wave experiments
  on chaotic scattering},\ }\href {https://doi.org/10.1088/0305-4470/38/49/001}
  {\bibfield  {journal} {\bibinfo  {journal} {J. Phys. A}\ }\textbf {\bibinfo
  {volume} {38}},\ \bibinfo {pages} {10433} (\bibinfo {year}
  {2005}{\natexlab{a}})}\BibitemShut {NoStop}%
\bibitem [{\citenamefont {Fyodorov}\ \emph {et~al.}(2005)\citenamefont
  {Fyodorov}, \citenamefont {Savin},\ and\ \citenamefont
  {Sommers}}]{fyodorov2005scattering}%
  \BibitemOpen
  \bibfield  {author} {\bibinfo {author} {\bibfnamefont {Y.~V.}\ \bibnamefont
  {Fyodorov}}, \bibinfo {author} {\bibfnamefont {D.}~\bibnamefont {Savin}},\
  and\ \bibinfo {author} {\bibfnamefont {H.}~\bibnamefont {Sommers}},\
  }\bibfield  {title} {\bibinfo {title} {Scattering, reflection and impedance
  of waves in chaotic and disordered systems with absorption},\ }\href
  {https://doi.org/10.1088/0305-4470/38/49/017} {\bibfield  {journal} {\bibinfo
   {journal} {J. Phys. A}\ }\textbf {\bibinfo {volume} {38}},\ \bibinfo {pages}
  {10731} (\bibinfo {year} {2005})}\BibitemShut {NoStop}%
\bibitem [{\citenamefont {Kuhl}\ \emph {et~al.}(2013)\citenamefont {Kuhl},
  \citenamefont {Legrand},\ and\ \citenamefont
  {Mortessagne}}]{kuhl2013microwave}%
  \BibitemOpen
  \bibfield  {author} {\bibinfo {author} {\bibfnamefont {U.}~\bibnamefont
  {Kuhl}}, \bibinfo {author} {\bibfnamefont {O.}~\bibnamefont {Legrand}},\ and\
  \bibinfo {author} {\bibfnamefont {F.}~\bibnamefont {Mortessagne}},\
  }\bibfield  {title} {\bibinfo {title} {{Microwave experiments using open
  chaotic cavities in the realm of the effective Hamiltonian formalism}},\
  }\href {https://doi.org/10.1002/prop.201200101} {\bibfield  {journal}
  {\bibinfo  {journal} {Fortschr. Phys.}\ }\textbf {\bibinfo {volume} {61}},\
  \bibinfo {pages} {404} (\bibinfo {year} {2013})}\BibitemShut {NoStop}%
\bibitem [{\citenamefont {Engelbrecht}\ and\ \citenamefont
  {Weidenm{\"u}ller}(1973)}]{engelbrecht1973hauser}%
  \BibitemOpen
  \bibfield  {author} {\bibinfo {author} {\bibfnamefont {C.}~\bibnamefont
  {Engelbrecht}}\ and\ \bibinfo {author} {\bibfnamefont {H.~A.}\ \bibnamefont
  {Weidenm{\"u}ller}},\ }\bibfield  {title} {\bibinfo {title} {{Hauser-Feshbach
  theory and Ericson fluctuations in the presence of direct reactions}},\
  }\href {https://doi.org/10.1103/PhysRevC.8.859} {\bibfield  {journal}
  {\bibinfo  {journal} {Phys. Rev. C}\ }\textbf {\bibinfo {volume} {8}},\
  \bibinfo {pages} {859} (\bibinfo {year} {1973})}\BibitemShut {NoStop}%
\bibitem [{\citenamefont {Hemmady}\ \emph
  {et~al.}(2005{\natexlab{a}})\citenamefont {Hemmady}, \citenamefont {Zheng},
  \citenamefont {Ott}, \citenamefont {Antonsen},\ and\ \citenamefont
  {Anlage}}]{hemmady2005universalPRL}%
  \BibitemOpen
  \bibfield  {author} {\bibinfo {author} {\bibfnamefont {S.}~\bibnamefont
  {Hemmady}}, \bibinfo {author} {\bibfnamefont {X.}~\bibnamefont {Zheng}},
  \bibinfo {author} {\bibfnamefont {E.}~\bibnamefont {Ott}}, \bibinfo {author}
  {\bibfnamefont {T.~M.}\ \bibnamefont {Antonsen}},\ and\ \bibinfo {author}
  {\bibfnamefont {S.~M.}\ \bibnamefont {Anlage}},\ }\bibfield  {title}
  {\bibinfo {title} {Universal impedance fluctuations in wave chaotic
  systems},\ }\href {https://doi.org/10.1103/PhysRevLett.94.014102} {\bibfield
  {journal} {\bibinfo  {journal} {Phys. Rev. Lett.}\ }\textbf {\bibinfo
  {volume} {94}},\ \bibinfo {pages} {014102} (\bibinfo {year}
  {2005}{\natexlab{a}})}\BibitemShut {NoStop}%
\bibitem [{\citenamefont {Hemmady}\ \emph
  {et~al.}(2005{\natexlab{b}})\citenamefont {Hemmady}, \citenamefont {Zheng},
  \citenamefont {Antonsen~Jr}, \citenamefont {Ott},\ and\ \citenamefont
  {Anlage}}]{hemmady2005universalPRE}%
  \BibitemOpen
  \bibfield  {author} {\bibinfo {author} {\bibfnamefont {S.}~\bibnamefont
  {Hemmady}}, \bibinfo {author} {\bibfnamefont {X.}~\bibnamefont {Zheng}},
  \bibinfo {author} {\bibfnamefont {T.~M.}\ \bibnamefont {Antonsen~Jr}},
  \bibinfo {author} {\bibfnamefont {E.}~\bibnamefont {Ott}},\ and\ \bibinfo
  {author} {\bibfnamefont {S.~M.}\ \bibnamefont {Anlage}},\ }\bibfield  {title}
  {\bibinfo {title} {Universal statistics of the scattering coefficient of
  chaotic microwave cavities},\ }\href
  {https://doi.org/10.1103/PhysRevE.71.056215} {\bibfield  {journal} {\bibinfo
  {journal} {Phys. Rev. E}\ }\textbf {\bibinfo {volume} {71}},\ \bibinfo
  {pages} {056215} (\bibinfo {year} {2005}{\natexlab{b}})}\BibitemShut
  {NoStop}%
\bibitem [{\citenamefont {Brouwer}(1995)}]{brouwer1995generalized}%
  \BibitemOpen
  \bibfield  {author} {\bibinfo {author} {\bibfnamefont {P.}~\bibnamefont
  {Brouwer}},\ }\bibfield  {title} {\bibinfo {title} {Generalized circular
  ensemble of scattering matrices for a chaotic cavity with nonideal leads},\
  }\href {https://doi.org/10.1103/PhysRevB.51.16878} {\bibfield  {journal}
  {\bibinfo  {journal} {Phys. Rev. B}\ }\textbf {\bibinfo {volume} {51}},\
  \bibinfo {pages} {16878} (\bibinfo {year} {1995})}\BibitemShut {NoStop}%
\bibitem [{\citenamefont {Kuhl}\ \emph
  {et~al.}(2005{\natexlab{b}})\citenamefont {Kuhl}, \citenamefont
  {Mart{\'\i}nez-Mares}, \citenamefont {M{\'e}ndez-S{\'a}nchez},\ and\
  \citenamefont {St{\"o}ckmann}}]{kuhl2005direct}%
  \BibitemOpen
  \bibfield  {author} {\bibinfo {author} {\bibfnamefont {U.}~\bibnamefont
  {Kuhl}}, \bibinfo {author} {\bibfnamefont {M.}~\bibnamefont
  {Mart{\'\i}nez-Mares}}, \bibinfo {author} {\bibfnamefont {R.}~\bibnamefont
  {M{\'e}ndez-S{\'a}nchez}},\ and\ \bibinfo {author} {\bibfnamefont {H.-J.}\
  \bibnamefont {St{\"o}ckmann}},\ }\bibfield  {title} {\bibinfo {title} {Direct
  processes in chaotic microwave cavities in the presence of absorption},\
  }\href {https://doi.org/10.1103/PhysRevLett.94.144101} {\bibfield  {journal}
  {\bibinfo  {journal} {Phys. Rev. Lett.}\ }\textbf {\bibinfo {volume} {94}},\
  \bibinfo {pages} {144101} (\bibinfo {year} {2005}{\natexlab{b}})}\BibitemShut
  {NoStop}%
\bibitem [{\citenamefont {Fyodorov}\ and\ \citenamefont
  {Savin}(2004)}]{fyodorov2004statistics}%
  \BibitemOpen
  \bibfield  {author} {\bibinfo {author} {\bibfnamefont {Y.~V.}\ \bibnamefont
  {Fyodorov}}\ and\ \bibinfo {author} {\bibfnamefont {D.~V.}\ \bibnamefont
  {Savin}},\ }\bibfield  {title} {\bibinfo {title} {Statistics of impedance,
  local density of states, and reflection in quantum chaotic systems with
  absorption},\ }\href {https://doi.org/10.1134/1.1868794} {\bibfield
  {journal} {\bibinfo  {journal} {J. Exp. Theor. Phys.}\ }\textbf {\bibinfo
  {volume} {80}},\ \bibinfo {pages} {725} (\bibinfo {year} {2004})}\BibitemShut
  {NoStop}%
\bibitem [{\citenamefont {Baranger}\ and\ \citenamefont
  {Mello}(1996)}]{baranger1996short}%
  \BibitemOpen
  \bibfield  {author} {\bibinfo {author} {\bibfnamefont {H.}~\bibnamefont
  {Baranger}}\ and\ \bibinfo {author} {\bibfnamefont {P.}~\bibnamefont
  {Mello}},\ }\bibfield  {title} {\bibinfo {title} {Short paths and information
  theory in quantum chaotic scattering: transport through quantum dots},\
  }\href {https://doi.org/10.1209/epl/i1996-00364-5} {\bibfield  {journal}
  {\bibinfo  {journal} {EPL}\ }\textbf {\bibinfo {volume} {33}},\ \bibinfo
  {pages} {465} (\bibinfo {year} {1996})}\BibitemShut {NoStop}%
\bibitem [{\citenamefont {Bulgakov}\ \emph {et~al.}(2006)\citenamefont
  {Bulgakov}, \citenamefont {Gopar}, \citenamefont {Mello},\ and\ \citenamefont
  {Rotter}}]{bulgakov2006statistical}%
  \BibitemOpen
  \bibfield  {author} {\bibinfo {author} {\bibfnamefont {E.~N.}\ \bibnamefont
  {Bulgakov}}, \bibinfo {author} {\bibfnamefont {V.~A.}\ \bibnamefont {Gopar}},
  \bibinfo {author} {\bibfnamefont {P.~A.}\ \bibnamefont {Mello}},\ and\
  \bibinfo {author} {\bibfnamefont {I.}~\bibnamefont {Rotter}},\ }\bibfield
  {title} {\bibinfo {title} {Statistical study of the conductance and shot
  noise in open quantum-chaotic cavities: Contribution from whispering gallery
  modes},\ }\href {https://doi.org/10.1103/PhysRevB.73.155302} {\bibfield
  {journal} {\bibinfo  {journal} {Phys. Rev. B}\ }\textbf {\bibinfo {volume}
  {73}},\ \bibinfo {pages} {155302} (\bibinfo {year} {2006})}\BibitemShut
  {NoStop}%
\bibitem [{\citenamefont {Hart}\ \emph {et~al.}(2009)\citenamefont {Hart},
  \citenamefont {Antonsen~Jr},\ and\ \citenamefont {Ott}}]{hart2009effect}%
  \BibitemOpen
  \bibfield  {author} {\bibinfo {author} {\bibfnamefont {J.~A.}\ \bibnamefont
  {Hart}}, \bibinfo {author} {\bibfnamefont {T.}~\bibnamefont {Antonsen~Jr}},\
  and\ \bibinfo {author} {\bibfnamefont {E.}~\bibnamefont {Ott}},\ }\bibfield
  {title} {\bibinfo {title} {Effect of short ray trajectories on the scattering
  statistics of wave chaotic systems},\ }\href
  {https://doi.org/10.1103/PhysRevE.80.041109} {\bibfield  {journal} {\bibinfo
  {journal} {Phys. Rev. E}\ }\textbf {\bibinfo {volume} {80}},\ \bibinfo
  {pages} {041109} (\bibinfo {year} {2009})}\BibitemShut {NoStop}%
\bibitem [{\citenamefont {Yeh}\ \emph {et~al.}(2010{\natexlab{a}})\citenamefont
  {Yeh}, \citenamefont {Hart}, \citenamefont {Bradshaw}, \citenamefont
  {Antonsen}, \citenamefont {Ott},\ and\ \citenamefont
  {Anlage}}]{yeh2010universal}%
  \BibitemOpen
  \bibfield  {author} {\bibinfo {author} {\bibfnamefont {J.-H.}\ \bibnamefont
  {Yeh}}, \bibinfo {author} {\bibfnamefont {J.~A.}\ \bibnamefont {Hart}},
  \bibinfo {author} {\bibfnamefont {E.}~\bibnamefont {Bradshaw}}, \bibinfo
  {author} {\bibfnamefont {T.~M.}\ \bibnamefont {Antonsen}}, \bibinfo {author}
  {\bibfnamefont {E.}~\bibnamefont {Ott}},\ and\ \bibinfo {author}
  {\bibfnamefont {S.~M.}\ \bibnamefont {Anlage}},\ }\bibfield  {title}
  {\bibinfo {title} {Universal and nonuniversal properties of wave-chaotic
  scattering systems},\ }\href {https://doi.org/10.1103/PhysRevE.81.025201}
  {\bibfield  {journal} {\bibinfo  {journal} {Phys. Rev. E}\ }\textbf {\bibinfo
  {volume} {81}},\ \bibinfo {pages} {025201} (\bibinfo {year}
  {2010}{\natexlab{a}})}\BibitemShut {NoStop}%
\bibitem [{\citenamefont {Yeh}\ \emph {et~al.}(2010{\natexlab{b}})\citenamefont
  {Yeh}, \citenamefont {Hart}, \citenamefont {Bradshaw}, \citenamefont
  {Antonsen}, \citenamefont {Ott},\ and\ \citenamefont
  {Anlage}}]{yeh2010experimental}%
  \BibitemOpen
  \bibfield  {author} {\bibinfo {author} {\bibfnamefont {J.-H.}\ \bibnamefont
  {Yeh}}, \bibinfo {author} {\bibfnamefont {J.~A.}\ \bibnamefont {Hart}},
  \bibinfo {author} {\bibfnamefont {E.}~\bibnamefont {Bradshaw}}, \bibinfo
  {author} {\bibfnamefont {T.~M.}\ \bibnamefont {Antonsen}}, \bibinfo {author}
  {\bibfnamefont {E.}~\bibnamefont {Ott}},\ and\ \bibinfo {author}
  {\bibfnamefont {S.~M.}\ \bibnamefont {Anlage}},\ }\bibfield  {title}
  {\bibinfo {title} {Experimental examination of the effect of short ray
  trajectories in two-port wave-chaotic scattering systems},\ }\href
  {https://doi.org/10.1103/PhysRevE.82.041114} {\bibfield  {journal} {\bibinfo
  {journal} {Phys. Rev. E}\ }\textbf {\bibinfo {volume} {82}},\ \bibinfo
  {pages} {041114} (\bibinfo {year} {2010}{\natexlab{b}})}\BibitemShut
  {NoStop}%
\bibitem [{\citenamefont {Savin}\ \emph {et~al.}(2017)\citenamefont {Savin},
  \citenamefont {Richter}, \citenamefont {Kuhl}, \citenamefont {Legrand},\ and\
  \citenamefont {Mortessagne}}]{savin2017fluctuations}%
  \BibitemOpen
  \bibfield  {author} {\bibinfo {author} {\bibfnamefont {D.~V.}\ \bibnamefont
  {Savin}}, \bibinfo {author} {\bibfnamefont {M.}~\bibnamefont {Richter}},
  \bibinfo {author} {\bibfnamefont {U.}~\bibnamefont {Kuhl}}, \bibinfo {author}
  {\bibfnamefont {O.}~\bibnamefont {Legrand}},\ and\ \bibinfo {author}
  {\bibfnamefont {F.}~\bibnamefont {Mortessagne}},\ }\bibfield  {title}
  {\bibinfo {title} {Fluctuations in an established transmission in the
  presence of a complex environment},\ }\href
  {https://doi.org/10.1103/PhysRevE.96.032221} {\bibfield  {journal} {\bibinfo
  {journal} {Phys. Rev. E}\ }\textbf {\bibinfo {volume} {96}},\ \bibinfo
  {pages} {032221} (\bibinfo {year} {2017})}\BibitemShut {NoStop}%
\bibitem [{\citenamefont {Savin}(2020)}]{savin2020statistics}%
  \BibitemOpen
  \bibfield  {author} {\bibinfo {author} {\bibfnamefont {D.~V.}\ \bibnamefont
  {Savin}},\ }\bibfield  {title} {\bibinfo {title} {Statistics of a simple
  transmission mode on a lossy chaotic background},\ }\href
  {https://doi.org/10.1103/PhysRevResearch.2.013246} {\bibfield  {journal}
  {\bibinfo  {journal} {Phys. Rev. Research}\ }\textbf {\bibinfo {volume}
  {2}},\ \bibinfo {pages} {013246} (\bibinfo {year} {2020})}\BibitemShut
  {NoStop}%
\bibitem [{\citenamefont {Haake}\ \emph {et~al.}(2018)\citenamefont {Haake},
  \citenamefont {Gnutzmann},\ and\ \citenamefont {Ku\'{s}}}]{haake_chapter}%
  \BibitemOpen
  \bibfield  {author} {\bibinfo {author} {\bibfnamefont {F.}~\bibnamefont
  {Haake}}, \bibinfo {author} {\bibfnamefont {S.}~\bibnamefont {Gnutzmann}},\
  and\ \bibinfo {author} {\bibfnamefont {M.}~\bibnamefont {Ku\'{s}}},\
  }\bibinfo {title} {{Quantum Signatures of Chaos}}\ (\bibinfo  {publisher}
  {Springer Series in Synergetics},\ \bibinfo {year} {2018})\BibitemShut
  {NoStop}%
\bibitem [{\citenamefont {Dupr{\'e}}\ \emph {et~al.}(2015)\citenamefont
  {Dupr{\'e}}, \citenamefont {del Hougne}, \citenamefont {Fink}, \citenamefont
  {Lemoult},\ and\ \citenamefont {Lerosey}}]{dupre2015wave}%
  \BibitemOpen
  \bibfield  {author} {\bibinfo {author} {\bibfnamefont {M.}~\bibnamefont
  {Dupr{\'e}}}, \bibinfo {author} {\bibfnamefont {P.}~\bibnamefont {del
  Hougne}}, \bibinfo {author} {\bibfnamefont {M.}~\bibnamefont {Fink}},
  \bibinfo {author} {\bibfnamefont {F.}~\bibnamefont {Lemoult}},\ and\ \bibinfo
  {author} {\bibfnamefont {G.}~\bibnamefont {Lerosey}},\ }\bibfield  {title}
  {\bibinfo {title} {{Wave-Field Shaping in Cavities: Waves Trapped in a Box
  with Controllable Boundaries}},\ }\href
  {https://doi.org/10.1103/PhysRevLett.115.017701} {\bibfield  {journal}
  {\bibinfo  {journal} {Phys. Rev. Lett.}\ }\textbf {\bibinfo {volume} {115}},\
  \bibinfo {pages} {017701} (\bibinfo {year} {2015})}\BibitemShut {NoStop}%
\bibitem [{\citenamefont {Gros}\ \emph {et~al.}(2020)\citenamefont {Gros},
  \citenamefont {del Hougne},\ and\ \citenamefont {Lerosey}}]{gros2019wave}%
  \BibitemOpen
  \bibfield  {author} {\bibinfo {author} {\bibfnamefont {J.-B.}\ \bibnamefont
  {Gros}}, \bibinfo {author} {\bibfnamefont {P.}~\bibnamefont {del Hougne}},\
  and\ \bibinfo {author} {\bibfnamefont {G.}~\bibnamefont {Lerosey}},\
  }\bibfield  {title} {\bibinfo {title} {Tuning a regular cavity to wave chaos
  with metasurface-reconfigurable walls},\ }\href
  {https://doi.org/10.1103/PhysRevA.101.061801} {\bibfield  {journal} {\bibinfo
   {journal} {Phys. Rev. A}\ }\textbf {\bibinfo {volume} {101}},\ \bibinfo
  {pages} {061801(R)} (\bibinfo {year} {2020})}\BibitemShut {NoStop}%
\bibitem [{\citenamefont {Verbaarschot}\ \emph {et~al.}(1985)\citenamefont
  {Verbaarschot}, \citenamefont {Weidenm{\"u}ller},\ and\ \citenamefont
  {Zirnbauer}}]{verbaarschot1985grassmann}%
  \BibitemOpen
  \bibfield  {author} {\bibinfo {author} {\bibfnamefont {J.}~\bibnamefont
  {Verbaarschot}}, \bibinfo {author} {\bibfnamefont {H.~A.}\ \bibnamefont
  {Weidenm{\"u}ller}},\ and\ \bibinfo {author} {\bibfnamefont {M.~R.}\
  \bibnamefont {Zirnbauer}},\ }\bibfield  {title} {\bibinfo {title} {Grassmann
  integration in stochastic quantum physics: The case of compound-nucleus
  scattering},\ }\href {https://doi.org/10.1016/0370-1573(85)90070-5}
  {\bibfield  {journal} {\bibinfo  {journal} {Phys. Rep.}\ }\textbf {\bibinfo
  {volume} {129}},\ \bibinfo {pages} {367} (\bibinfo {year}
  {1985})}\BibitemShut {NoStop}%
\bibitem [{\citenamefont {Mitchell}\ \emph {et~al.}(2010)\citenamefont
  {Mitchell}, \citenamefont {Richter},\ and\ \citenamefont
  {Weidenm{\"u}ller}}]{mitchell2010random}%
  \BibitemOpen
  \bibfield  {author} {\bibinfo {author} {\bibfnamefont {G.~E.}\ \bibnamefont
  {Mitchell}}, \bibinfo {author} {\bibfnamefont {A.}~\bibnamefont {Richter}},\
  and\ \bibinfo {author} {\bibfnamefont {H.~A.}\ \bibnamefont
  {Weidenm{\"u}ller}},\ }\bibfield  {title} {\bibinfo {title} {Random matrices
  and chaos in nuclear physics: Nuclear reactions},\ }\href
  {https://doi.org/10.1103/RevModPhys.82.2845} {\bibfield  {journal} {\bibinfo
  {journal} {Rev. Mod. Phys.}\ }\textbf {\bibinfo {volume} {82}},\ \bibinfo
  {pages} {2845} (\bibinfo {year} {2010})}\BibitemShut {NoStop}%
\bibitem [{\citenamefont {Fyodorov}\ and\ \citenamefont
  {Savin}(2011)}]{dima_book_chapter}%
  \BibitemOpen
  \bibfield  {author} {\bibinfo {author} {\bibfnamefont {Y.~V.}\ \bibnamefont
  {Fyodorov}}\ and\ \bibinfo {author} {\bibfnamefont {D.~V.}\ \bibnamefont
  {Savin}},\ }\bibfield  {title} {\bibinfo {title} {Resonance scattering of
  waves in chaotic systems},\ }in\ \href@noop {} {\emph {\bibinfo {booktitle}
  {The Oxford Handbook of Random Matrix Theory}}},\ \bibinfo {editor} {edited
  by\ \bibinfo {editor} {\bibfnamefont {G.}~\bibnamefont {Akemann}}, \bibinfo
  {editor} {\bibfnamefont {J.}~\bibnamefont {Baik}},\ and\ \bibinfo {editor}
  {\bibfnamefont {P.}~\bibnamefont {Di~Francesco}}}\ (\bibinfo  {publisher}
  {Oxford University Press, UK},\ \bibinfo {year} {2011})\ Chap.~\bibinfo
  {chapter} {34}, pp.\ \bibinfo {pages} {703--722},\ \bibinfo {note}
  {[arXiv:1003.0702]}\BibitemShut {NoStop}%
\bibitem [{\citenamefont {Poli}\ \emph {et~al.}(2009)\citenamefont {Poli},
  \citenamefont {Savin}, \citenamefont {Legrand},\ and\ \citenamefont
  {Mortessagne}}]{poli2009statistics}%
  \BibitemOpen
  \bibfield  {author} {\bibinfo {author} {\bibfnamefont {C.}~\bibnamefont
  {Poli}}, \bibinfo {author} {\bibfnamefont {D.~V.}\ \bibnamefont {Savin}},
  \bibinfo {author} {\bibfnamefont {O.}~\bibnamefont {Legrand}},\ and\ \bibinfo
  {author} {\bibfnamefont {F.}~\bibnamefont {Mortessagne}},\ }\bibfield
  {title} {\bibinfo {title} {Statistics of resonance states in open chaotic
  systems: A perturbative approach},\ }\href
  {https://doi.org/10.1103/PhysRevE.80.046203} {\bibfield  {journal} {\bibinfo
  {journal} {Phys. Rev. E}\ }\textbf {\bibinfo {volume} {80}},\ \bibinfo
  {pages} {046203} (\bibinfo {year} {2009})}\BibitemShut {NoStop}%
\bibitem [{Note1()}]{Note1}%
  \BibitemOpen
  \bibinfo {note} {Optimizing $H^0$ rather than $V$ has the inconvenience of
  resulting in $N^2+1$ as opposed to $NM+1$ parameters to be
  optimized.}\BibitemShut {Stop}%
\bibitem [{\citenamefont {Kraft}(1988)}]{kraft1988software}%
  \BibitemOpen
  \bibfield  {author} {\bibinfo {author} {\bibfnamefont {D.}~\bibnamefont
  {Kraft}},\ }\href@noop {} {\bibinfo {title} {{A Software Package for
  Sequential Quadratic Programming}}} (\bibinfo {year} {1988})\BibitemShut
  {NoStop}%
\bibitem [{\citenamefont {Savin}(2018)}]{savin2018envelope}%
  \BibitemOpen
  \bibfield  {author} {\bibinfo {author} {\bibfnamefont {D.~V.}\ \bibnamefont
  {Savin}},\ }\bibfield  {title} {\bibinfo {title} {Envelope and phase
  distribution of a resonance transmission through a complex environment},\
  }\href {https://doi.org/10.1103/PhysRevE.97.062202} {\bibfield  {journal}
  {\bibinfo  {journal} {Phys. Rev. E}\ }\textbf {\bibinfo {volume} {97}},\
  \bibinfo {pages} {062202} (\bibinfo {year} {2018})}\BibitemShut {NoStop}%
\bibitem [{\citenamefont {Holloway}\ \emph {et~al.}(2012)\citenamefont
  {Holloway}, \citenamefont {Shah}, \citenamefont {Pirkl}, \citenamefont
  {Young}, \citenamefont {Hill},\ and\ \citenamefont
  {Ladbury}}]{holloway2012reverberation}%
  \BibitemOpen
  \bibfield  {author} {\bibinfo {author} {\bibfnamefont {C.~L.}\ \bibnamefont
  {Holloway}}, \bibinfo {author} {\bibfnamefont {H.~A.}\ \bibnamefont {Shah}},
  \bibinfo {author} {\bibfnamefont {R.~J.}\ \bibnamefont {Pirkl}}, \bibinfo
  {author} {\bibfnamefont {W.~F.}\ \bibnamefont {Young}}, \bibinfo {author}
  {\bibfnamefont {D.~A.}\ \bibnamefont {Hill}},\ and\ \bibinfo {author}
  {\bibfnamefont {J.}~\bibnamefont {Ladbury}},\ }\bibfield  {title} {\bibinfo
  {title} {Reverberation chamber techniques for determining the radiation and
  total efficiency of antennas},\ }\href
  {https://doi.org/10.1109/TAP.2012.2186263} {\bibfield  {journal} {\bibinfo
  {journal} {IEEE Trans. Antennas Propag.}\ }\textbf {\bibinfo {volume} {60}},\
  \bibinfo {pages} {1758} (\bibinfo {year} {2012})}\BibitemShut {NoStop}%
\bibitem [{\citenamefont {Gros}\ and\ \citenamefont
  {Lerosey}(2018)}]{conferenceJB}%
  \BibitemOpen
  \bibfield  {author} {\bibinfo {author} {\bibfnamefont {J.-B.}\ \bibnamefont
  {Gros}}\ and\ \bibinfo {author} {\bibfnamefont {G.}~\bibnamefont {Lerosey}},\
  }\bibfield  {title} {\bibinfo {title} {Random matrix model of wave front
  shaping in chaotic cavities},\ }\href
  {http://dx.doi.org/10.13140/RG.2.2.10459.59689} {\bibfield  {journal}
  {\bibinfo  {journal} {Complex2018 Summer School: Transport, Mesoscopy and
  Imaging of Waves in Complex Media}\ } (\bibinfo {year} {2018})}\BibitemShut
  {NoStop}%
\bibitem [{\citenamefont {K{\"o}ber}\ \emph {et~al.}(2010)\citenamefont
  {K{\"o}ber}, \citenamefont {Kuhl}, \citenamefont {St{\"o}ckmann},
  \citenamefont {Gorin}, \citenamefont {Savin},\ and\ \citenamefont
  {Seligman}}]{kober2010microwave}%
  \BibitemOpen
  \bibfield  {author} {\bibinfo {author} {\bibfnamefont {B.}~\bibnamefont
  {K{\"o}ber}}, \bibinfo {author} {\bibfnamefont {U.}~\bibnamefont {Kuhl}},
  \bibinfo {author} {\bibfnamefont {H.-J.}\ \bibnamefont {St{\"o}ckmann}},
  \bibinfo {author} {\bibfnamefont {T.}~\bibnamefont {Gorin}}, \bibinfo
  {author} {\bibfnamefont {D.}~\bibnamefont {Savin}},\ and\ \bibinfo {author}
  {\bibfnamefont {T.}~\bibnamefont {Seligman}},\ }\bibfield  {title} {\bibinfo
  {title} {Microwave fidelity studies by varying antenna coupling},\ }\href
  {https://doi.org/10.1103/PhysRevE.82.036207} {\bibfield  {journal} {\bibinfo
  {journal} {Phys. Rev. E}\ }\textbf {\bibinfo {volume} {82}},\ \bibinfo
  {pages} {036207} (\bibinfo {year} {2010})}\BibitemShut {NoStop}%
\bibitem [{\citenamefont {Kumar}(2015)}]{kumar2015random}%
  \BibitemOpen
  \bibfield  {author} {\bibinfo {author} {\bibfnamefont {S.}~\bibnamefont
  {Kumar}},\ }\bibfield  {title} {\bibinfo {title} {{Random matrix ensembles
  involving Gaussian Wigner and Wishart matrices, and biorthogonal
  structure}},\ }\href {https://doi.org/10.1103/PhysRevE.92.032903} {\bibfield
  {journal} {\bibinfo  {journal} {Phys. Rev. E}\ }\textbf {\bibinfo {volume}
  {92}},\ \bibinfo {pages} {032903} (\bibinfo {year} {2015})}\BibitemShut
  {NoStop}%
\bibitem [{Note2()}]{Note2}%
  \BibitemOpen
  \bibinfo {note} {See the Supplemental Material for details on the alternative
  parametric model.}\BibitemShut {Stop}%
\bibitem [{\citenamefont {Davy}\ \emph {et~al.}(2016)\citenamefont {Davy},
  \citenamefont {de~Rosny},\ and\ \citenamefont {Besnier}}]{davy2016green}%
  \BibitemOpen
  \bibfield  {author} {\bibinfo {author} {\bibfnamefont {M.}~\bibnamefont
  {Davy}}, \bibinfo {author} {\bibfnamefont {J.}~\bibnamefont {de~Rosny}},\
  and\ \bibinfo {author} {\bibfnamefont {P.}~\bibnamefont {Besnier}},\
  }\bibfield  {title} {\bibinfo {title} {Greenâ€™s function retrieval with
  absorbing probes in reverberating cavities},\ }\href
  {https://doi.org/10.1103/PhysRevLett.116.213902} {\bibfield  {journal}
  {\bibinfo  {journal} {Phys. Rev. Lett.}\ }\textbf {\bibinfo {volume} {116}},\
  \bibinfo {pages} {213902} (\bibinfo {year} {2016})}\BibitemShut {NoStop}%
\end{thebibliography}
\end{document}

\centerline{\bf SUPPLEMENTAL MATERIAL:  }
\centerline{\bf Implementing Non-Universal Features with a Random Matrix Theory Approach:  }
\centerline{\bf Application to Space-to-Configuration Multiplexing  }
\bigskip

\centerline{Philipp del Hougne,\textsuperscript{1} Dmitry V. Savin,\textsuperscript{2} Olivier Legrand,\textsuperscript{1} and Ulrich Kuhl\textsuperscript{1}}

\bigskip

\centerline{\textsuperscript{1} \small{\textit{Institut de Physique de Nice, CNRS UMR 7010, Universit\'{e} C\^ote d'Azur, 06108 Nice, France}}}
\centerline{\textsuperscript{2} \small{\textit{Department of Mathematics, Brunel University London, Uxbridge, UB8 3PH, United Kingdom}}}

\bigskip
\bigskip
\bigskip

For the interested reader, here we provide additional details on the use of an alternative parametric model with $H=H^0 + \lambda CC^\dagger$. 
This model could be motivated by a treatment of each metasurface element as an active scattering center or ``channel'' in one of two possible distinct reflecting states~\cite{conferenceJB}, which can be described by a low-rank perturbation of that type~\cite{kober2010microwave}. However, as summarized in the main text, this model has two important drawbacks compared to the $ H = \textrm{cos}(\lambda)H^0 + \textrm{sin}(\lambda) H^s$ model used in the main text: (i) it is significantly more complicated to evaluate, and (ii) its performance in terms of reproducing the distributions of the experimentally measured scattering parameters and effective ranks is inferior. Nonetheless, for completeness, we summarize the results that we obtained with this alternative model in this Supplemental Material.

\subsection{Overview of alternative parametric model}

The alternative parametric model that we investigate in this Supplemental Material is

\begin{equation}\label{Eq_S}
     S^{RMT}(E) = I_M - i V^T \frac{1}{ \left( E+i\frac{\Gamma_a}{2}\right)I_N - \left( H^0 + \lambda C C^\dagger  -\frac{i}{2} V V^T \right) } V ,
\end{equation}

\noindent where $I_\alpha$ denotes the $\alpha\times\alpha$ identity matrix, $H^0$ is a $N \times N$ matrix drawn from the GOE, $\Gamma_a$ represents the average mode width, $V$ is the $N \times M$ coupling matrix, and $C$ is a $N \times N_{pix}$ matrix. $C$ is generated in the same way that the coupling matrix $V$ of a traditional RMT model would be generated for a system with $N_{pix}$ ports with all $\kappa_i=1$. As in the main text, we choose $N=250$ in sight of the computational cost of optimizing $V$. $N_{pix}$ is the (effective) number of independent metasurface pixels and we define the variable $\beta = N_{pix}/N$. The variable $\lambda$ weights stirred and unstirred component of the Hamiltonian.

\subsection{Impact of $\lambda$ and $\beta$ on the eigenvalue PDF of the Hamiltonian}

\begin{figure} [h] \centering
\includegraphics [width = 6cm]{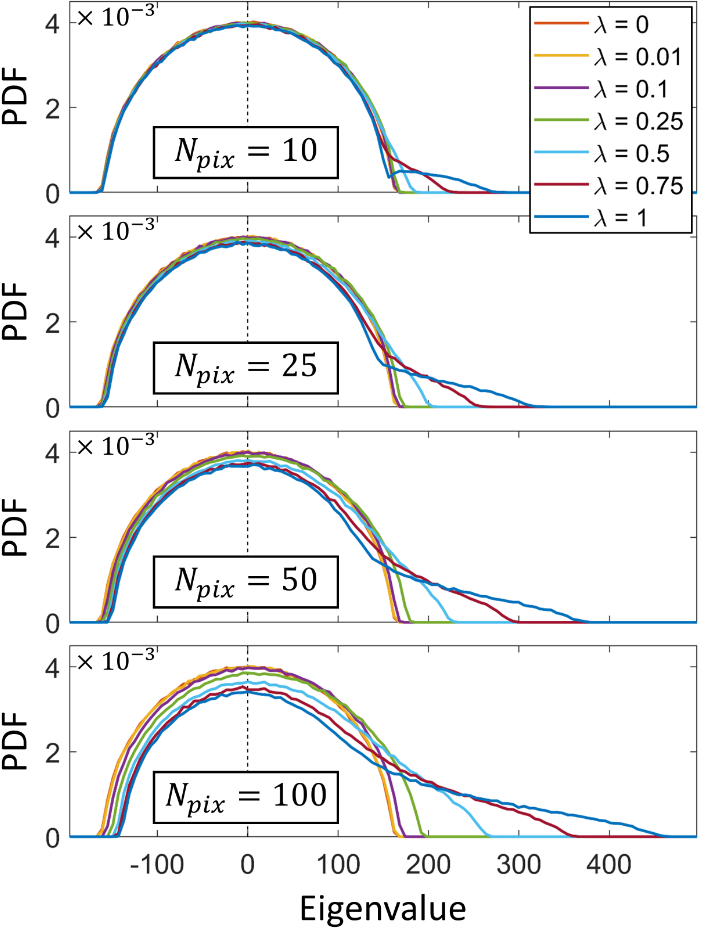}
\caption{Eigenvalue PDF of Hamiltonian for alternative parametric model with different values of $\beta=N_{pix}/N$ and $\lambda$.}
\label{figS1}
\end{figure}

Since the diagonal entries of $C$ are real and positive, they alter the eigenvalue PDF of the Hamiltonian $H=H^0 + \lambda C C^\dagger$. We begin by plotting the eigenvalue PDF for a range of different values of $\lambda$ for four values of $\beta$ in Fig.~\ref{figS1}.
The results suggest that the position of the eigenvalue PDF's peak does not appear to be impacted by the addition of the stirred Hamiltonian. However, the maximum value of the eigenvalue PDF is reduced.

We continue to work at $E=0$ as in the main text. To counteract the lower value of the PDF of eigenvalues at $E=0$, we consider replacing $\Gamma_a$ by $\Gamma_a^\prime = \alpha \Gamma_a$, where $\alpha$ is the ratio of the slope of the curve of sorted eigenvalues around zero for $H$ and $H^0$. 
However, since we find that accounting for the change in slope or not does not have an appreciable impact on the final results, we neglect the correction in the following. 
We choose $N_{pix}=50$ in the following which ensures $N_{pix}<N$ to preserve the GOE shape of the level density near the origin and $N_{pix} \gg 1$ to reflect the fact that we have a multitude of pixels rather than a single or a few pixels.

\subsection{Optimal value of $\lambda$ and resulting $\langle S^{RMT} \rangle$}

\begin{figure}
\centering
\includegraphics [width = 246pt]{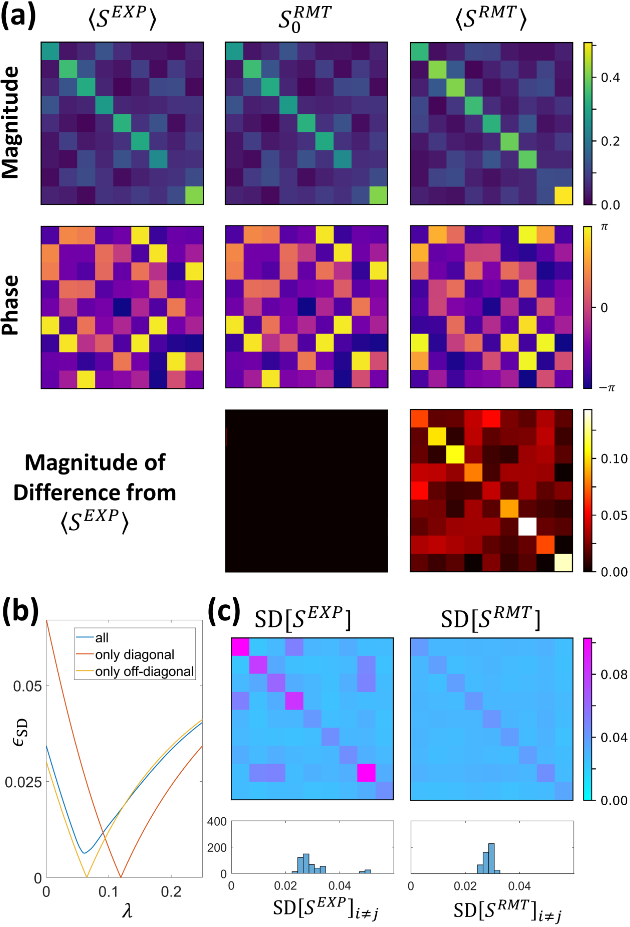}
\caption{Counterpart for the alternative parametric model to Fig.~2 from the main text. (a) Comparison of $\langle S^{EXP} \rangle$, $S^{RMT}_0$ and $\langle S^{RMT} \rangle$ in terms of magnitude and phase. The magnitude of the difference of the latter two from $\langle S^{EXP} \rangle$ is also shown. (b) Dependence of $\epsilon_{SD}$ on $\lambda$ considering all entries, only the diagonal entries or only the off-diagonal entries of the scattering matrix. (c) SD of $S^{EXP}$ and $S^{RMT}$ and PDF of the SD of the off-diagonal scattering matrix entries.}
\label{figS2}
\end{figure}

Since the optimization of $V$ is assumed to only dependent on $H^0$ in our two-step optimization procedure, it is the same as for the parametric model discussed in the main text; therefore, we only discuss the optimization of $\lambda$ for the alternative parametric model here. We present in Fig.~\ref{figS2} the counterpart to Fig.~2 from the main text. The optimal value of $\lambda \approx 0.06$ is comparable to that found with the parametric model in the main text. 
However, $\langle S^{RMT}\rangle$ does not resemble $\langle S^{EXP}\rangle$ as much as for the parametric model presented in the main text. In Fig.~\ref{figS2}(a), the differences of the magnitude can be spotted by the naked eye and the magnitude of the difference between $\langle S^{RMT}\rangle$ and $\langle S^{EXP}\rangle$ is an order of magnitude larger. The alternative model is thus not as suitable for our two-step optimization procedure (which assumes $S^{RMT}_0=\langle S^{RMT} \rangle$ is true to very good approximation) as the model from the main text. In terms of the SDs of the scattering matrix entries shown in Fig.~\ref{figS2}(c), no notable difference from Fig.~2(c) is seen.

\clearpage

\subsection{$R_{\mathrm{eff}}(\mathcal{H})$ distribution}

\begin{figure}
\centering
\includegraphics [width = 460pt]{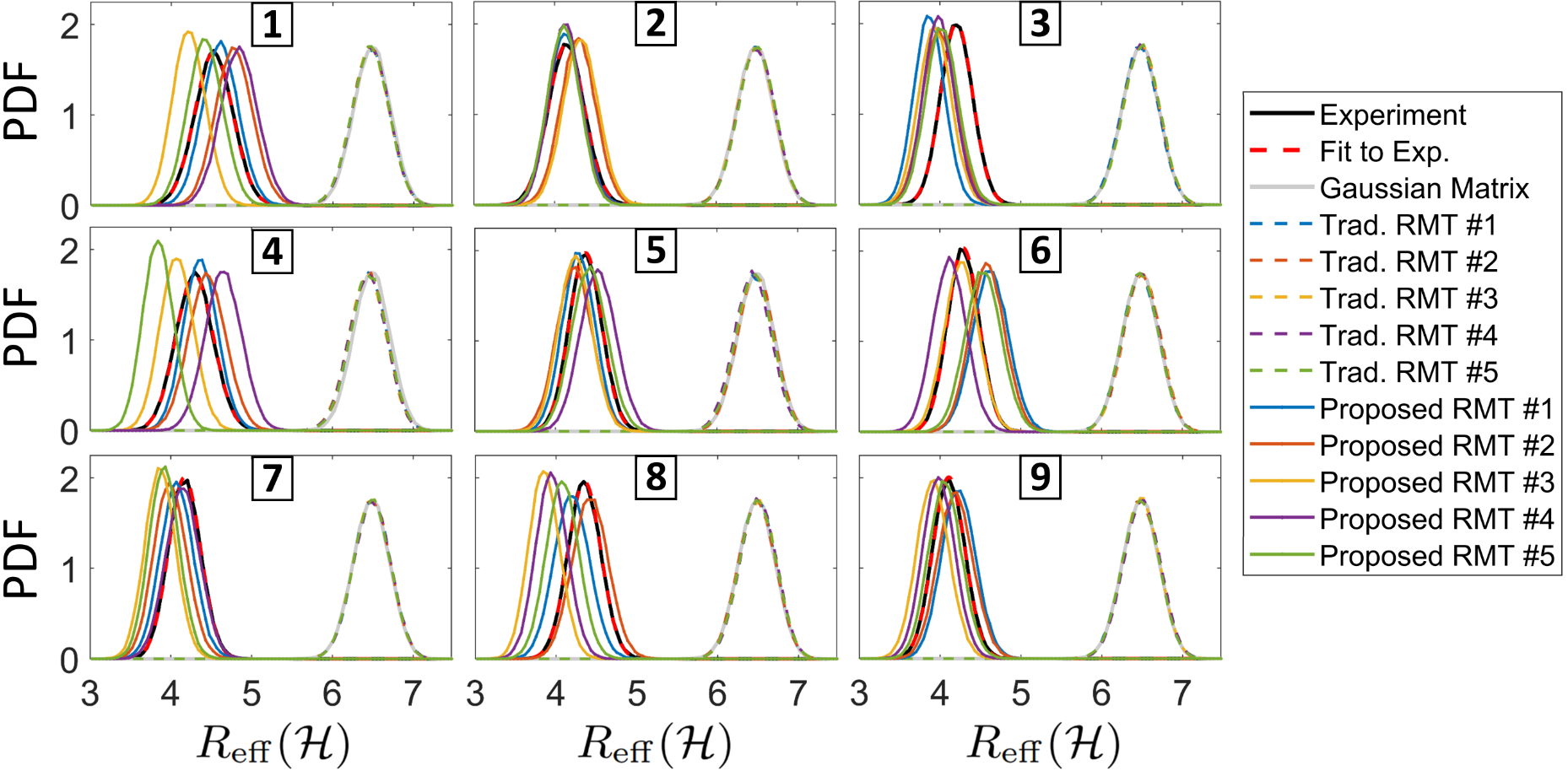}
\caption{Counterpart for the alternative parametric model to Fig.~3 from the main text. For the nine different possible choices of receive port (number indicated as inset), we compare the PDF of $R_{\mathrm{eff}}(\mathcal{H})$ extracted from $S^{EXP}$ with that extract from $S^{RMT}$. For the latter, five different optimization runs (starting with a different random $H^0$) are shown. A Gaussian fit to the experimental distribution is also shown. Moreover, we show the PDFs for five different ensembles with traditional RMT as well as for a simple complex Gaussian matrix with i.i.d. entries.}
\label{figS3}
\end{figure}

Consequently, in comparison to Fig.~3 in the main text, the distributions of $R_{\mathrm{eff}}(\mathcal{H})$ predicted by the alternative parametric model, displayed in Fig.~\ref{figS3}, are visibly in  inferior agreement with the experimentally observed ones. Moreover, the dependence on the specific optimization run (starting with a different random $H^0$) is seen to be stronger.

\clearpage

\subsection{$S_{i,j}$ distribution}

\begin{figure} 
\centering
\includegraphics [width = 246pt]{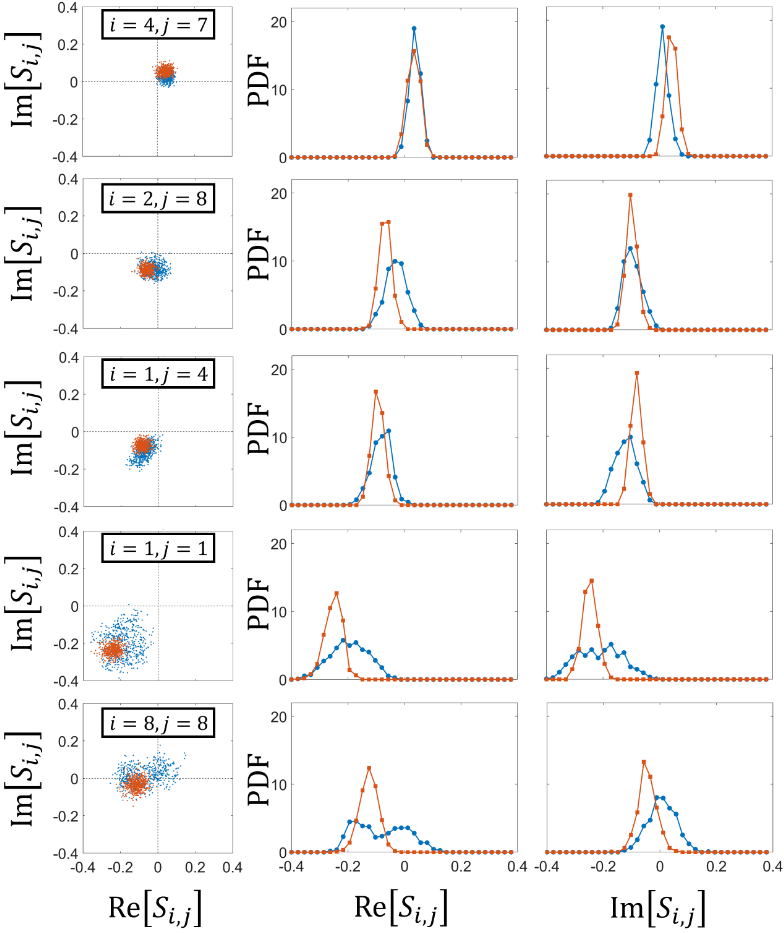}
\caption{Counterpart for the alternative parametric model to Fig.~4 from the main text. Distribution of selected $S_{i,j}$ in the complex plane and corresponding PDFs of $\mathrm{Re}[S_{i,j}]$ and $\mathrm{Im}[S_{i,j}]$. Experiment and simulation are presented in blue and red, respectively.}
\label{figS4}
\end{figure}

For completeness, we finally provide the counterpart to Fig.~4 from the main text in Fig.~\ref{figS4}. We note that even for scattering matrix entries like $S_{1,4}$ for which the SD is correctly captured and which are correctly reproduced by the model discussed in the main text, the center of the cloud generated by the alternative model is dislocated. This is essentially due to the inferior match in terms of $\langle S \rangle$ already seen in Fig.~\ref{figS2}(a).

\newpage

\providecommand{\noopsort}[1]{}\providecommand{\singleletter}[1]{#1}%